\documentclass[11pt,a4paper]{article}

\usepackage{graphicx}
\usepackage{dcolumn}
\usepackage{bm}
\usepackage{caption}
\usepackage{epsfig,multicol,bbm}
\usepackage{amsmath,amssymb}
\usepackage[labelsep=colon]{caption}

\newcommand\fverb{\setbox\fverbbox=\hbox\bgroup\verb}
\newcommand\fverbdo{\egroup\medskip\noindent%
            \fbox{\unhbox\fverbbox}\ }
\newcommand\fverbit{\egroup\item[\fbox{\unhbox\fverbbox}]}
\newbox\fverbbox

\begin{document}

\title{{$\beta$-decay properties of
waiting point nuclei for astrophysical applications}
\author{Jameel-Un Nabi$^{*}$, Munir Ahmad and Gul Daraz}\\}
 \maketitle

\begin{abstract}
We report microscopic calculation of key $\beta$-decay properties
for some of the crucial  waiting point species having neutron closed
magic shells 50 and 82. Our calculation bear astrophysical
significance vis-\'{a}-vis speeding of the $r$-process. The
$\beta$-decay  properties include half-lives, energy rates of
$\beta$-delayed neutrons and their emission probabilities, both
under terrestrial and stellar conditions. We perform a pn-QRPA
calculation with a separable multi-shell interaction and include
both allowed and unique first-forbidden (U1F) transitions in our
calculation. We compare our results with previous calculations and
measured data.  Our calculation is in good agreement with the
experimental data. For certain cases, we noted a significant
decrease in the half-life calculation with the contribution of U1F
transitions. This is in contradiction to the shell model study where
only for $N$ = 126 waiting-point nuclei, the forbidden transitions
were reported to significantly reduce the calculated half-lives. Our
model fulfills the Ikeda sum rule for even-even cases. For odd-A
cases the rule is violated by up to 15$\%$ for $^{125}$Tc.
\begin{description}
\item[Keywords]{$\beta$-decay half-lives, pn-QRPA, Gamow-Teller transitions,
U1F transitions, Ikeda sum rule, $r$-process.}
\item[PACS number(s)]{23.40.2s, 26.30.1k, 97.60.Bw, 98.80.Ft}
\end{description}
\end{abstract}

\section{Introduction}
Since the seminal paper on synthesis of elements in stars
\cite{Bur57}, our understanding of the nucleosynthesis process have
greatly improved (for recent papers see e.g. \cite{Cyb16,Nom13}).
The $r$- and $s$-processes are the mastermind responsible for
nucleosynthesis of heavy elements. The $r$-process mechanism
basically requires the understanding of nuclear characteristics of
hundreds of neutron-rich nuclide, mostly unknown. The weak
interaction rates and reaction cross sections are amongst the key
nuclear input data to affect the $r$-process calculation. The
$\beta$-decay rates deserve a special mention as they are
responsible for changing the nuclear specie during heavy element
synthesis. At the same time $r$-process mechanism also demands
accurate estimate of other physical parameters including entropy,
temperature, density and lepton-to-baryon ratio of the stellar
matter. The physical conditions conducive for occurrence of
$r$-process are relatively high temperatures (of the order of few
GK) and high neutron densities ($>$ 10$^{20}$ cm$^{-3}$)
\cite{Bur57,Cow91,Arn99,Kra93,Woo94}. Under prevailing conditions
the capturing process of neutrons takes place at a faster pace than
the competing $\beta$-decay processes and many neutron-rich nuclei
(with $S_{n}\lesssim 3$ MeV) are produced. Nuclei possessing closed
neutron shells of 50, 82 and 126  exhibit discontinuities in neutron
separation energies because of stronger binding energy. Consequently
the $r$-process matter flow decelerates and wait for occurrence of
several $\beta$-decays to occur before the process of rapid neutron
capture resumes. Peaks have been observed in the distribution
abundances of $r$- mechanism at $N$ = 50, 82 and 126 because of
accumulation of matter at these waiting point nuclei. The calculated
half-lives of $\beta$-decay for waiting point nuclei describes the
time scale it takes the mass flow to transpose seed nuclei to larger
nuclei in the third peak at around $A \sim$ 200. Provided that the
$r$- mechanism has enough duration time for $\beta$-flow equilibrium
to built, the $\beta$-decay half-lives are proportional to the
relative elemental abundances \cite{Kra88}.

Unfortunately the experimental information for waiting point nuclei
is scarce. For the neutron closed shells of $N$ = 50 and 82 waiting
point nuclei, the available half-lives are rather limited and
insufficient \cite{Hos05,Kra98,Kra86,Pfe01}. The scenario is
expected to improve in near future with radioactive ion beam
experiments at RIKEN \cite{Nis11} and GSI \cite{Kur09}. Hence for
$r$-process simulations the required $\beta$-decay half-lives come
primarily from theoretical estimations. An extensive tabulation of
microscopic $\beta$-decay rates for a wide range of nuclei was
reported by  Klapdor-Kleingrothaus et al. \cite{Kla84}. Later,
Staudt et al. \cite{Sta89,Sta90} and Hirsh et al. \cite{Hir93}, used
the proton-neutron quasiparticle random phase approximation
(pn-QRPA) model, for the first time, to predict $\beta$-decay
half-lives for a wide range of proton-rich and neutron-rich exotic
nuclei. The $r$-process spectra of waiting point nuclei can be
affected by the presence of low lying energy levels possessing
different parities. This necessitates the incorporation of  the
first-forbidden (FF) chapter to the $\beta$-decay half-lives. The
pn-QRPA model was used to estimate the FF contributions for a
handful of nuclei for the first time by Homma et al. \cite{Hom96}.
Later other models were used to estimate the FF contribution. These
include, but are not limited to, the QRPA + gross theory
\cite{Mol03}, self-consistent density-functional + continuum QRPA
\cite{Bor05} and more recently the large-scale shell model
calculation \cite{Zhi13}. Only a small percent of the total $3
(N-Z)$ sum rule lie within the $Q_{\beta}$ window for the
neutron-rich nuclei participating in the $r$-process. The rest of
the strength resides in the Gamow-Teller (GT) giant resonance
located at much larger excitation energies. This may furnish
explanation as to why different model calculations of $\beta$-decay
half-lives may differ significantly without violating the sum rule.

Taking the weak-interaction rates to stellar domain is a next level
calculation.   Nabi and Klapdor-Kleingrothaus used the pn-QRPA
approach and calculated stellar weak rates of $sd-$, $fp-$ and
$fpg$-shell nuclei \cite{Nab99d,Nab99c,Nab04} for various
astrophysical applications. The current pn-QRPA approach, using a
separable interaction with a multi-$\hbar\omega$ space, makes
possible a state-by-state calculation of weak interaction rates
summing over Boltzmann-weighted, microscopically estimated GT
strengths for all parent excited levels. This distinguishing feature
of current calculation makes it unique amongst all calculations of
stellar weak rates (including those using the independent particle
model and shell model).

In the present work we report the GT strength distribution
calculation, half-life calculation, stellar $\beta$-decay and
positron emission rate calculations, energy rates of $\beta$-delayed
neutron and corresponding neutron emission probabilities ($P_{n}$)
for nuclei having neutron magic numbers ($N$ = 50 and 82)  using the
pn-QRPA model. Ten waiting point nuclei (six having $N$= 50 and four
having $N$ = 82) were selected for this paper. In all cases we
consider both the allowed GT and unique first-forbidden (U1F)
transition contribution to the total weak rates. Non-unique
transitions are also important. Currently we are working on codes to
calculate non-unique contributions and their inclusion would be
taken as a future assignment. We organize our paper in four
sections. Section~2 describes the necessary pn-QRPA formalism. In
Section~3, we show our results and present comparison with
measurement and previous calculations. Conclusions and our key
findings are drawn in Section~4.

\section{Theoretical Formalism}
The addition of all transition probabilities to levels in the
daughter state $j$  with energies $E_{j}$ lying within the
$Q_{\beta}$ window gives the terrestrial half-life of $\beta$-decay.
\begin{equation}\label{T1}
T_{1/2}=({\sum_{0\leqslant E_{j}\leqslant Q_{\beta}}
{1/}{t_{j}}})^{-1},
\end{equation}
where $t_{j}$ shows the  partial half-life for the allowed
 $\beta$-decay transition given by
\begin{equation}\label{f_0}
f_{0}(Z,Q_{\beta}-E_{j})t_{j}=\frac{D}{(g_{A}/g_{v})^2 B(E_{j})},
\end{equation}
here ($g_{A}$/$g_{v}$) is axial to vector coupling constant ratio,
(numerical value is -1.254), $D$ is a physical constant  given by
$D$=$2{\pi^{3}}\hbar^{7}\ln2/{g^{2}_v}{m^{5}_e}{c^{4}}$ (numerical
value is 6295 s) and $f_{0}$ is the Fermi integral function (taking
into account finite size effects and screening of nucleus, using the
recipe of Gove and Martin \cite{Gov71}). The $B(E_{j})$ gives the
reduced transition probabilities (including GT and Fermi
transitions). Our model includes GT force with separable
particle-hole ($ph$) and particle-particle ($pp$) matrix elements.
The two forces were characterized by strength parameter $\chi_{GT}$
and $\kappa_{GT}$, respectively. For the pn-QRPA Hamiltonian, its
model parameters, and calculation of reduced transition
probabilities, we refer to \cite{Sta90, Hir93}. The formalism is not
repeated here for space consideration.

For the U1F transitions the $pp$ and $ph$ matrix elements are given
by
\begin{equation}
\label{vph} V^{ph}_{pn,p^{\prime}n^{\prime}} = +2\chi_{U1F}
f_{pn}(\mu)f_{p^{\prime}n^{\prime}}(\mu),
\end{equation}

\begin{equation}\label{vpp}
V^{pp}_{pn,p^{\prime}n^{\prime}} = -2\kappa_{U1F}
f_{pn}(\mu)f_{p^{\prime}n^{\prime}}(\mu),
\end{equation}

where
\begin{equation}\label{fpn}
f_{pn}(\mu)=<p|t_{-}r[\sigma Y_{1}]_{2\mu}|n>,
\end{equation}
is a single particle transition amplitude between Nilsson single
particle states (deformed). Here $\mu$ values are labeled as,
$\mu=0,\pm1$ and $\pm2$ (and represents the spherical component of
the transition operator). Other symbols have regular meaning. The
neutron and proton states possess different parities \cite{Hom96}.

We varied the $ph$ and $pp$ strength interaction constant within the
specified limits (ensuring the QRPA calculation does not
"collapses"). Guidelines for choosing the interaction constants were
taken from \cite{Hom96,Nab16}. The idea was to come up with an
analytical formula for $\chi$ and $\kappa$ that best reproduced the
measured half-lives within 1$\sigma$ deviation. Measured half-life
data were taken from \cite{Aud12}. We obtained the following mass
dependent relationship for these constants:

$\chi_{GT}$ =  $64.6/A$ MeV; $\chi_{U1F}$ = $64.6/A$ MeV fm$^{-2}$, \\
$\kappa_{GT}$ = $5.6/A$ MeV; $\kappa_{U1F}$ = $11.7/A$ MeV
fm$^{-2}$.

The deformation parameter was taken from the \cite{Moe95}, while
Q-values were taken from \cite{Aud12}.

The stellar $\beta$-decay rates for allowed GT and U1F transitions
from parent ($\mathit{i}$th level) to daughter ($\mathit{j}$th
level) nucleus were determined using
\begin{equation}
\label{lij} \lambda_{ij}^{\beta} =
\frac{m_{e}^{5}c^{4}}{2\pi^{3}\hbar^{7}}\sum_{\Delta
J^{\pi}}g^{2}f_{ij}(\Delta J^{\pi})B_{ij}(\Delta J^{\pi}),
\end{equation}
in the above equation $B_{ij}(\Delta J^{\pi})$ and $f_{ij}(\Delta
J^{\pi})$ are the reduced transition probability and phase space
factor respectively. For allowed transitions  the reduced GT
($\Delta J^{\pi}$ =1$^{+}$) transition probabilities are given by
\begin{equation}\label{bgt}
B(GT)_{ij} = \frac{1}{2J_{i}+1} \mid <j
\parallel \sum_{k}t_{-}^{k}\vec{\sigma}^{k} \parallel i> \mid ^{2}.
\end{equation}
The reduced Fermi ($\Delta J^{\pi}$ =0$^{+}$) transition
probabilities are given by
\begin{equation}
B(F)_{ij} = \frac{1}{2J_{i}+1} \mid<j \parallel \sum_{k}t_{-}^{k}
\parallel i> \mid ^{2}.
\end{equation}

The phase space integral $(f_{ij})$ is an integral over total
energy. For the case of $\beta$-decay it is given by (from here
onwards we use natural units, $\hbar=m_{e}=c=1$).
\begin{equation}\label{ps}
f_{ij} = \int_{1}^{w_{m}} w \sqrt{w^{2}-1} (w_{m}-w)^{2} F(+ Z,w)
(1-G_-) dw,
\end{equation}
whereas for continuum positron capture phase space is given by
\begin{equation}\label{pc}
f_{ij} = \int_{w_{l}}^{\infty} w \sqrt{w^{2}-1} (w_{m}+w)^{2} F(-
Z,w) G_+ dw,
\end{equation}

For the U1F transitions,

\begin{eqnarray}\label{e}
B_{ij}(\Delta
J^{\pi})=\frac{1}{12}z^{2}(w_{m}^{2}-1)-\frac{1}{6}z^{2}w_{m}w+\frac{1}{6}z^{2}w^{2},
\end{eqnarray}
where $z$ is
\begin{eqnarray}\label{efg}
z=2g_{A}\frac{\langle j|\vert\sum_{k}r_{k}[\textbf{C}^{k}_{1}\times
\boldsymbol{\sigma}]^{2}{\textbf{t}}^{k}_{-}\vert|i\rangle}{\sqrt{2J_{i}+1}},
\end{eqnarray}
\begin{eqnarray}\label{gkl}
\textbf{C}_{lm}=\sqrt{\frac{4\pi}{2l+1}}\textbf{Y}_{lm},
\end{eqnarray}
$\textbf{Y}_{lm}$ are the spherical harmonics. In case of U1F
interaction, $f_{ij}$ (phase space integral) were calculated using
\begin{eqnarray}\label{f}
f_{ij} = \int_{1}^{w_{m}} w \sqrt{w^{2}-1}
(w_{m}-w)^{2}[(w_{m}-w)^{2}F_{1}(Z,w) \nonumber\\
+ (w^{2}-1)F_{2}(Z,w)] (1-G_{-})dw,
\end{eqnarray}
where the upper limit of the integral gives the total $\beta$-decay
energy given by ($ w_{m} = m_{p}-m_{d}+E_{i}-E_{j}$,). $w$ is the
total energy of the electron including its rest mass. One should
note that if the corresponding electron emission total energy,
$w_{m}$, is greater than -1, then $w_{l}=1$, and if it is less than
or equal to 1, then $w_{l}=\mid w_{m} \mid$. The $G_{+}$ and $G_{-}$
are the positron and electron distribution functions, respectively.
The  $F(\pm Z,w)$ , $F_{1}(Z,w)$ and $F_{2}( Z,w)$ are the Fermi
functions computed using the recipe of \cite{Gov71}.

The high temperature inside the core of massive stars signifies that
there is a finite probability of occupancy of parent excited levels
in stellar scenario. Using the assumption of thermal equilibrium the
occupation probability of $\mathit{i}$th state can be computed using
\begin{equation}\label{pi}
P_{i} = \frac {exp(-E_{i}/kT)}{\sum_{i=1}exp(-E_{i}/kT)}.
\end{equation}

Finally  stellar $\beta$-decay rate per unit time per nucleus was
determined using
\begin{equation}
\label{lb} \lambda^{\beta} = \sum_{ij}P_{i} \lambda_{ij}^{\beta}.
\end{equation}
A similar sum was performed to calculate continuum positron capture
rates in stellar matter. Summations was carried out for all initial
states as well as for final states until desired convergence were
obtained in our rate calculation. In our calculation it was further
assumed that all
 daughter excited states having energy larger than the
neutron separation energy ($S_{n}$), decayed by neutron emission.
The energy rate for neutron emission from daughter system was
determined using
\begin{equation}\label{ln}
\lambda^{n} = \sum_{ij}P_{i}\lambda_{ij}(E_{j}-S_{n}),
\end{equation}
for all $E_{j} > S_{n}$. The probability of $\beta$-delayed neutron
emission, $P_{n}$, was calculated using
\begin{equation}\label{pn}
P_{n} =
\frac{\sum_{ij\prime}P_{i}\lambda_{ij\prime}}{\sum_{ij}P_{i}\lambda_{ij}},
\end{equation}
where $j\prime$ indicates the energy levels of the daughter nucleus
with $E_{j\prime}
> S_{n}$.
The $\lambda_{ij(\prime)}$ in Eq.~(\ref{ln}) and Eq.~(\ref{pn}),
represents the sum of positron capture and electron emission rates,
for transition arising from $i$ $\rightarrow$ $j(j\prime)$.

\section{Results and discussion}
In this section we are going to present the terrestrial
$\beta$-decay half-lives, stellar weak rates, phase space  and
charge-changing strength distribution calculations, including both
allowed GT and U1F transitions. The predictive power of the pn-QRPA
model becomes more effective for smaller $T_{1/2}$ values (with
increasing distance from the stability line) \cite{Hir93,Nab16}
which justifies the usage of present model for $\beta$-decay
calculations. We compare our calculation with several previous
pioneering calculations \cite{Mol03,Bor05,Zhi13,Pfe02,Mol97} as well
as against experimental data \cite{Aud12}. We multiplied results of
pn-QRPA calculated strength by a quenching factor of $f_{q}^{2}$ =
(0.55)$^{2}$ \cite{nab15b} in order to compare them with
experimental data and prior calculations, and to later use them in
astrophysical reaction rates.

The computed $\beta$-decay terrestrial half-lives for $r$-process
waiting point nuclei having $N$ = 50 and 82, including allowed GT
and U1F contributions, are shown in Table~\ref{tab1}. Here we also
show the shell model calculations \cite{Lan03,Cue07} with only
allowed GT contribution, the large scale shell model calculation
\cite{Zhi13} including both allowed GT and first-forbidden (FF)
contributions and the QRPA calculation performed by \cite{Mol03}
where the allowed GT part was calculated using the QRPA model and
gross theory was employed to calculate the FF contribution.
\begin{table}[h]
\begin{center}
\caption{Comparison of our computed $\beta$-decay half-lives for $N$
= 50 and 82 $r$-process waiting point nuclei with previous
calculations and experimental half-lives. Half-lives mentioned with
an asterisk in the last column were adopted from \cite{Mol03}.}
\label{tab1}
  \tiny
\begin{tabular}{cc|c|c|cc|cc|c}
    \multicolumn{2}{c}{} & \multicolumn{5}{c}{} \\
        \hline
    \multicolumn{2}{c|}{}  & {SM \cite{Lan03,Cue07}}  & {LSSM \cite{Zhi13}}  & \multicolumn{2}{c|}{QRPA+ Gross Theory \cite{Mol03}}  & \multicolumn{2}{c|}{This work} & Exp.\cite{Aud12a} \\
    \hline

    Nucleus    & A & $T_{1/2}$ & $T_{1/2}$ & $T_{1/2}$ & $T_{1/2}$ & $T_{1/2}$ & $T_{1/2}$ & \multicolumn{1}{c}{$T_{1/2}$}
    \\
         &  & (GT) & (GT+FF) & (GT) & (GT+FF) & (GT) & (GT+U1F) & \multicolumn{1}{c}{} \\
     \hline
    Fe         & 76    & 0.008 & 0.008 & 0.045 & 0.027 & 0.015 & 0.014 & \multicolumn{1}{c}{0.013$^{\ast}$} \\
    Co         & 77    & 0.016 & 0.016 & 0.013 & 0.014 & 0.013 & 0.010 & \multicolumn{1}{c}{0.010$^{\ast}$} \\
    Ni         & 78    & 0.127 & 0.150 & 0.477 & 0.224 & 0.210 & 0.152 & \multicolumn{1}{c}{0.140} \\
    Cu         & 79    & 0.222 & 0.270 & 0.430 & 0.157 & 0.273 & 0.239 & \multicolumn{1}{c}{0.220} \\
    Zn         & 80    & 0.432 & 0.530 & 3.068 & 1.260 & 0.910 & 0.634 & \multicolumn{1}{c}{0.550} \\
    Ga         & 81    & 0.577 & 1.030 & 1.568 & 1.227 & 1.509 & 1.457 & \multicolumn{1}{c}{1.217} \\
    Tc         & 125   & 0.009 & 0.010 & 0.009 & 0.009 & 0.031 & 0.008 & \multicolumn{1}{c}{0.008$^{\ast}$} \\
    Ru         & 126   & 0.020 & 0.020 & 0.034 & 0.030 & 0.709 & 0.019 & \multicolumn{1}{c}{0.017$^{\ast}$} \\
    Rh         & 127   & 0.028 & 0.028 & 0.022 & 0.020 & 0.109 & 0.074 & \multicolumn{1}{c}{0.070$^{\ast}$} \\
    Pd         & 128   & 0.046 & 0.047 & 0.125 & 0.074 & 2.431 & 0.025 & \multicolumn{1}{c}{0.020} \\
    \end{tabular}
    \label{T1}
\end{center}
\end{table}
\begin{table}[h]
\begin{center}
\caption{Comparison of our computed $\beta$-decay half-lives for $N$
= 50 $r$-process waiting point nuclei with previous calculations and
experimental half-lives. $a \rightarrow $ \cite{Bor05} and $b
\rightarrow$ \cite{Aud12a}.} \label{tab2}
  \tiny
    \begin{tabular}{cc|c|c|c|ccc|cc}

     \multicolumn{2}{c}{} & \multicolumn{8}{c}{} \\
\hline
    \multicolumn{2}{c|}{}  &Exp. &Borzov \cite{Bor05} & M\"{o}ller \cite{Mol97} & \multicolumn{3}{c|}{Pfeiffer et al. \cite{Pfe02}} & \multicolumn{2}{c}{This work} \\
    \hline
    Nuclei     & A     & $T_{1/2}$& $T_{1/2}$ &$T_{1/2}$  & $T_{1/2}$ & $T_{1/2}$ & $T_{1/2}$ & $T_{1/2}$ &
    $T_{1/2}$\\
     &      & &(GT+FF) &(GT)  &(KHF) & (QRPA-1) & (QRPA-2) & (GT) & (GT+U1F)\\
    \hline
    Co       & 77    & ---   &--- & ---&0.020 & 0.010 & 0.015 & 0.013 & 0.010 \\
    Ni       & 78    & 0.110${^a}$ &0.134 &0.489 & 0.066 & 0.332 & 0.326 & 0.210 & 0.152 \\
    Cu       & 79    & 0.220${^b}$ &0.182 &0.276 &0.076 & 0.358 & 0.212 & 0.273 & 0.239 \\
    Zn       & 80    & 0.550${^b}$ &1.039 &--- &0.255 & 3.025 & 2.033 & 0.910 & 0.634 \\
    Ga       & 81    & 1.217${^b}$ &1.322 &1.555 &0.404 & 1.684 & 1.852 & 1.509 & 1.457
    \\\\
    \end{tabular}
    \label{T2}
\end{center}
\end{table}
Experimental half-lives were taken from the recent available atomic
mass data evaluation of \cite{Aud12a}. Table~\ref{tab2} shows a
similar comparison of our calculated $\beta$-decay half-lives of $N$
= 50 $r$-process waiting point nuclei with previous QRPA
calculations and measured data. Here we compare with the
self-consistent density-functional and continuum QRPA framework
including the GT and FF transition calculation \cite{Bor05}, a QRPA
calculation using finite-range droplet model and folded-Yukawa
single particle potential \cite{Mol97} and the QRPA calculations by
\cite{Pfe02}. For details of KHF, QRPA-1 and QRPA-2 calculations we
refer to \cite{Pfe02}.

It may be seen from Table~\ref{tab1} and Table~\ref{tab2} that our
calculated $T_{1/2}$ values are in very good comparison with the
measured half-lives. Besides few $N$ = 50 nuclei, the U1F
contribution substantially lowers the calculated half-lives,
specially for $N$ = 82 cases.

The $\beta$-delayed neutron emission probabilities were also
estimated by employing the QRPA  \cite{Bor05,Pfe02} and the shell
model \cite{Zhi13} approaches. Table~\ref{tab3} compares our pn-QRPA
calculated $\beta$-delay neutron emission probabilities against
previous calculations and experimental predictions. Noticeable
differences between shell model and our calculated probabilities are
seen in Table~\ref{tab3}. Our numbers are in decent agrement with
the QRPA calculations of \cite{Pfe02}.

\begin{table}[h]
\begin{center}
\caption{Comparison between theoretical and experimental predictions
of  $\beta$-delayed neutron emission probability values for the
selected waiting point nuclei.} \label{tab3}
 \tiny
    \begin{tabular}{cc|c|c|cc|ccc|c}
    \multicolumn{2}{c}{} & \multicolumn{8}{c}{} \\
        \hline
   \multicolumn{2}{c|}{}  & Exp.\cite{Pfe02} & LSSM \cite{Zhi13} & \multicolumn{2}{c|}{ Borzov \cite{Bor05}}& \multicolumn{3}{c|}{Pfeiffer et al. \cite{Pfe02}} & This work \\
    \hline

Nucl.     & A     & $P_{n}$& $P_{n}$ &$P_{n}$  & $P_{n}$ &$P_{n}$
      &$P_{n}$ & $P_{n}$ &
    $P_{n}$\\
        &    & &  & (GT) & (GT+FF)  & (KHF) & (QRPA-1) & (QRPA-2) & (GT+U1F) \\
    \hline
    Co        & 77    & --  & 77.2 &--- & --- & 52.8  & 39.3  & 78.1  & 100.0 \\
    Ni        & 78    & --  & 79 & 51.4&51.0 &10.8  & 40.7  & 55.7  & 11.0 \\
    Cu        & 79    & 55  & 88.6 &64.8 &63.4 & 21.8  & 33.7  & 27.9  & 15.0 \\
    Zn        & 80    & 1   & 14.1 &3.8 &4.2 & 0.7   & 10.9  & 10.0  & 0.2 \\
    Ga        & 81    & 12.1 & 13 & 14.5 &17.1 & 3.8   & 6.7   & 7.0   & 63.0 \\

    \end{tabular}
    \label{T3}
\end{center}
\end{table}
\begin{table}[h]
\begin{center}
\caption{Statistical data of pn-QRPA calculated GT strength
distributions.} \label{tab4}
  \scriptsize
    \begin{tabular}{ccc|ccc|cc}

    \multicolumn{3}{c}{Waiting-point nuclei} &        \multicolumn{3}{c}{Gamow-Teller Data} & \multicolumn{2}{c}{Ikeda Sum Rule} \\
    \hline
    Nucleus  & Z     & A     & Centroid B(GT-)  & Width B(GT-) & $\sum$B(GT-) &  Calculated  & Theoretical \\
    \hline
    Fe    & 26    & 76    & 48.0  & 5.1   & 72.1     & 72.0  & 72 \\
    Co    & 27    & 77    & 49.8  & 5.6   & 69.8    & 68.9  & 69 \\
    Ni    & 28    & 78    & 44.4  & 4.0   & 66.0       & 66.0  & 66 \\
    Cu    & 29    & 79    & 49.3  & 4.8   & 62.9     & 62.8  & 63 \\
    Zn    & 30    & 80    & 39.1  & 4.0   & 60.7     & 60.0  & 60 \\
    Ga    & 31    & 81    & 40.8  & 3.5   & 56.7     & 56.6  & 57 \\
    Tc    & 43    & 125   & 48.4  & 2.9   & 99.2       & 99.2  & 117 \\
    Ru    & 44    & 126   & 41.6  & 3.4   & 114.0      & 114.0 & 114 \\
    Rh    & 45    & 127   & 45.9  & 3.2   & 97.9       & 97.9  & 111 \\
    Pd    & 46    & 128   & 38.8  & 2.9   & 108.4    & 108.0 & 108 \\
    \end{tabular}
    \label{T4}

\end{center}
\end{table}

The total GT strength (in $\beta$-decay direction), centroid and
width of calculated GT strength distributions for the $N$ = 50 and
82 waiting point nuclei using our pn-QRPA model are shown in
Table~\ref{tab4}. The table reveals placement of GT centroid at high
excitation energies in daughter nuclei. This necessitates the
calculation of charge-changing transitions up to high excitation
energies in daughter. Only a large model space (up to 7 major
shells) made this calculation in the present formalism possible.
Shown also in Table~\ref{tab4} is the comparison of our calculated
Ikeda sum rule with the theoretical prediction (which is model
independent). It may be seen that the Ikeda sum rule is fulfilled
for even-even cases. For odd-A cases the compliance is 85$\%$ and
88$\%$ for $^{125}$Tc and $^{127}$Rh, respectively. For remaining
odd-A cases the compliance is greater than 99$\%$.

\begin{figure}[h]
\includegraphics[width=3.8in]{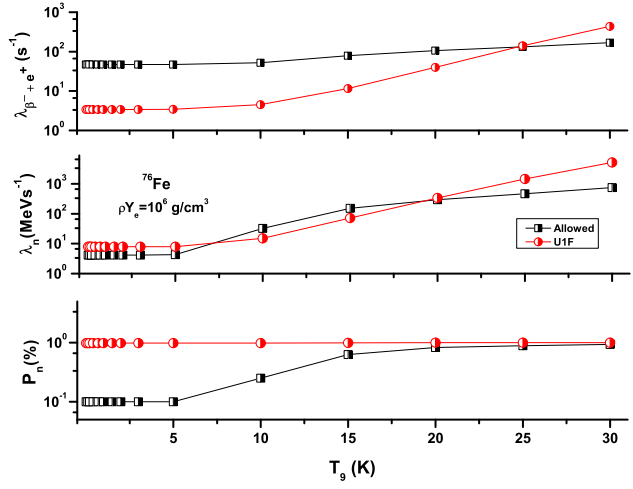}
\caption{\scriptsize The pn-QRPA calculated $\beta^{-}$ decay and
positron capture rates (upper panel), energy rates of
$\beta$-delayed neutron (middle panel) and their emission
probabilities (bottom panel) for $^{76}$Fe as a function of core
temperature at stellar density of 10$^{6}$g.cm$^{-3}$. The allowed
GT and U1F contributions are shown separately.}\label{figure1}
\centering
\end{figure}

\begin{figure}[t!]
\includegraphics[width=3.8in]{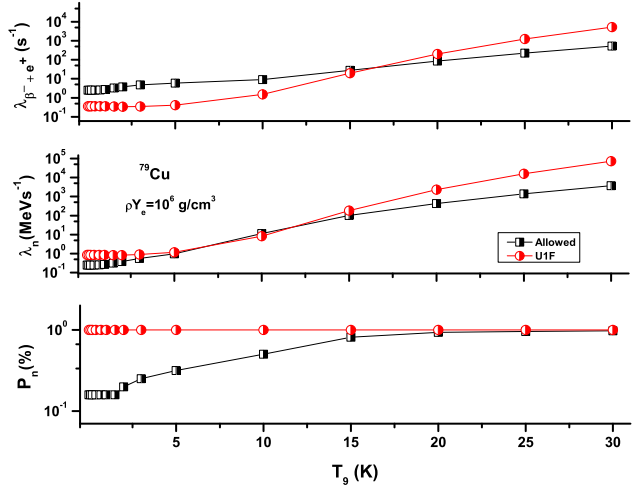}
\caption{\scriptsize Same as Fig.~\ref{figure1} but for
$^{79}$Cu.}\label{figure2} \centering
\end{figure}

\begin{figure}[t!]
\includegraphics[width=3.8in]{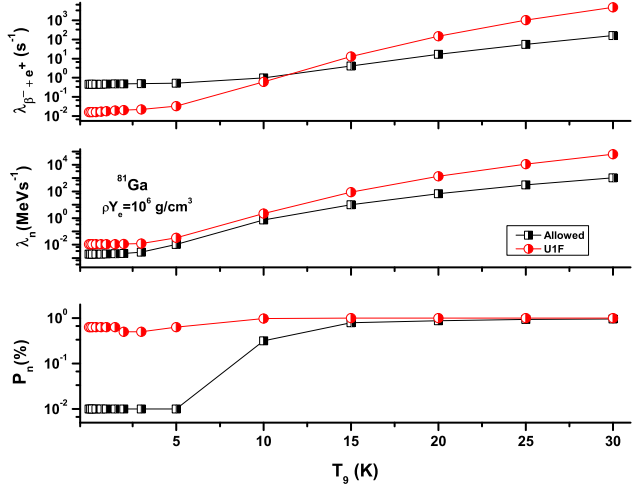}
\caption{\scriptsize Same as Fig.~\ref{figure1} but for
$^{81}$Ga.}\label{figure3} \centering
\end{figure}

\begin{figure}[t!]
\includegraphics[width=3.8in]{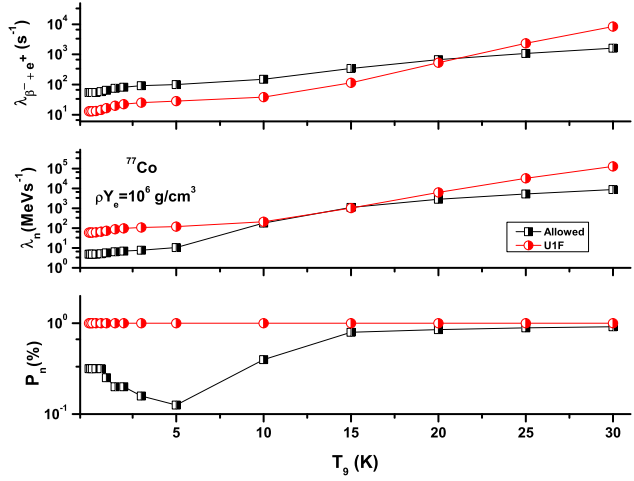}
\caption{\scriptsize Same as Fig.~\ref{figure1} but for
$^{77}$Co.}\label{figure4} \centering
\end{figure}

\begin{figure}[t!]
\includegraphics[width=3.8in]{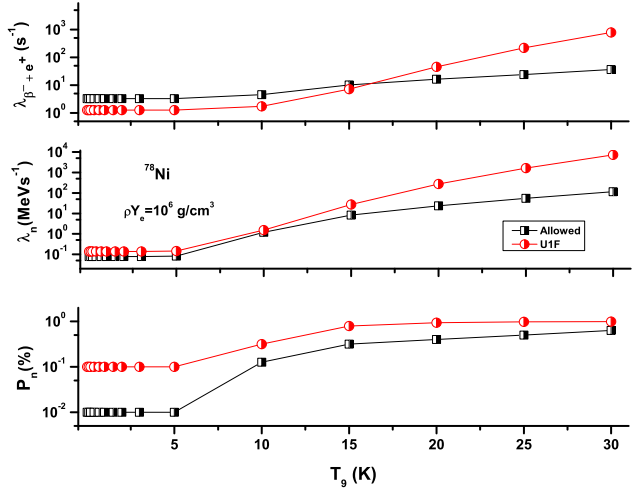}
\caption{\scriptsize Same as Fig.~\ref{figure1} but for
$^{78}$Ni.}\label{figure5} \centering
\end{figure}

\begin{figure}[t!]
\includegraphics[width=3.8in]{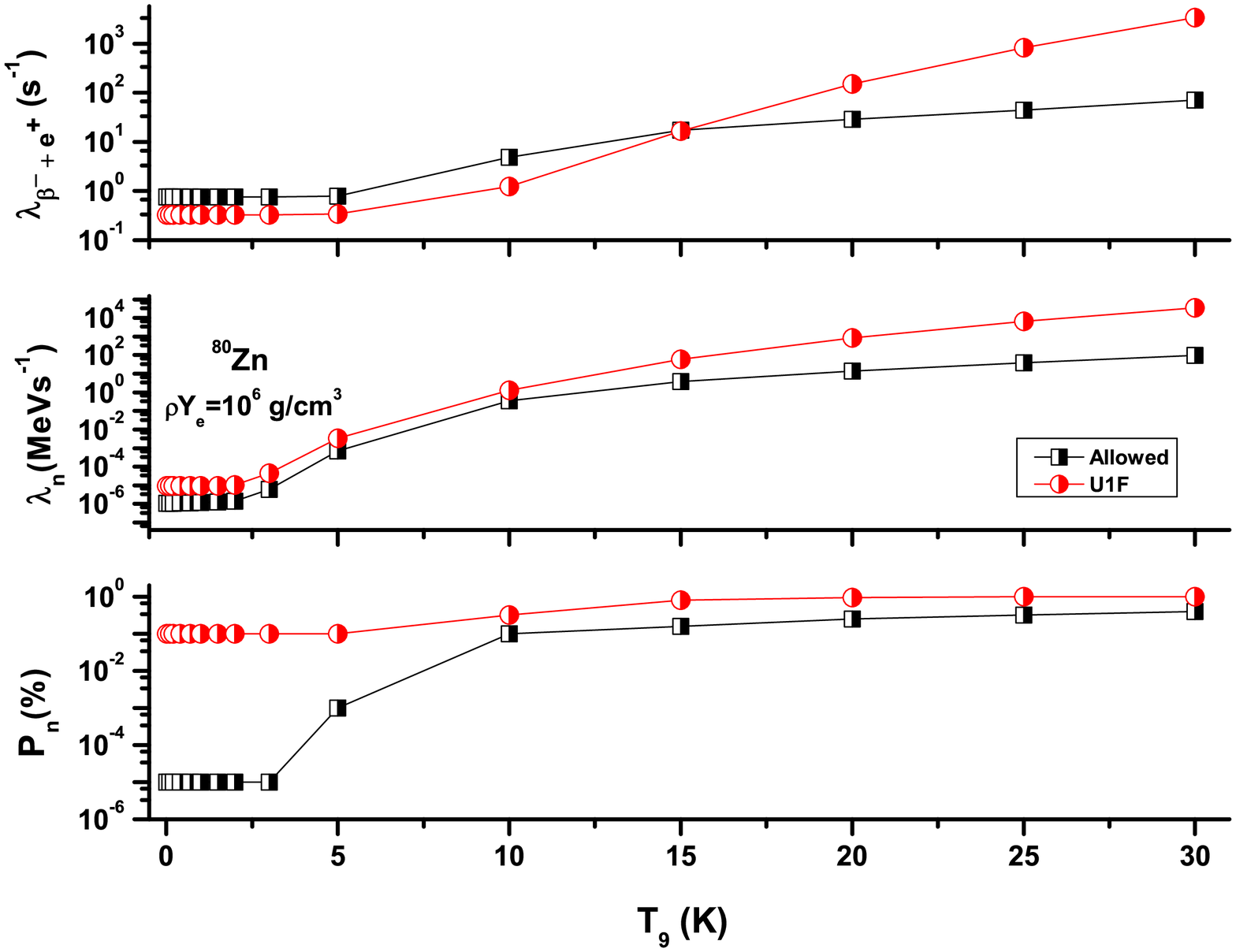}
\caption{\scriptsize Same as Fig.~\ref{figure1} but for
$^{80}$Zn.}\label{figure6} \centering
\end{figure}

\begin{figure}[t!]
\includegraphics[width=3.8in]{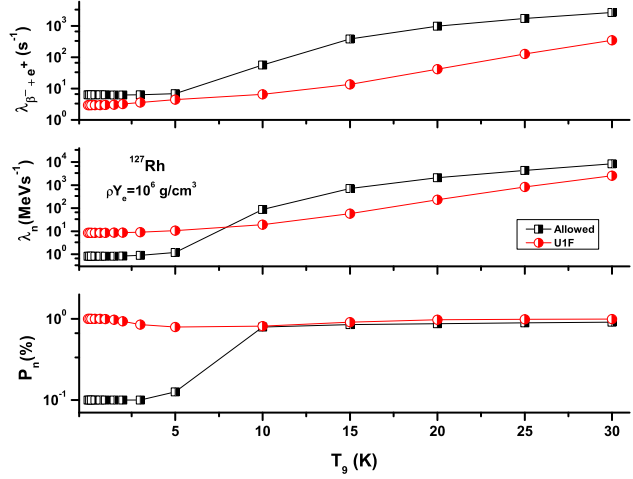}
\caption{\scriptsize Same as Fig.~\ref{figure1} but for
$^{127}$Rh.}\label{figure7} \centering
\end{figure}

\begin{figure}[t!]
\includegraphics[width=3.8in]{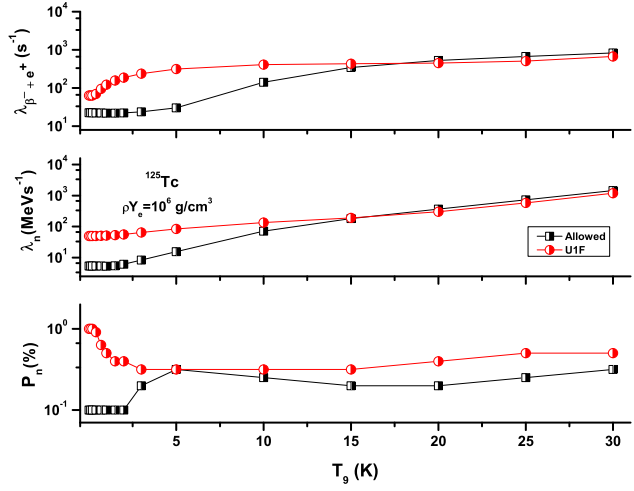}
\caption{\scriptsize Same as Fig.~\ref{figure1} but for
$^{125}$Tc.}\label{figure8} \centering
\end{figure}

\begin{figure}[t!]
\includegraphics[width=3.8in]{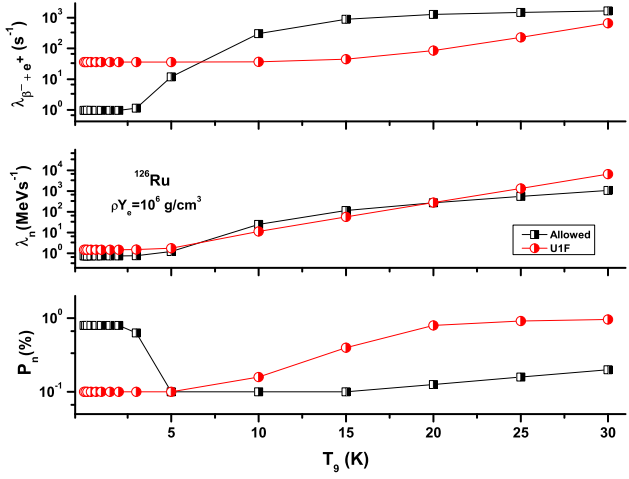}
\caption{\scriptsize Same as Fig.~\ref{figure1} but for
$^{126}$Ru.}\label{figure9} \centering
\end{figure}

\begin{figure}[t!]
\includegraphics[width=3.8in]{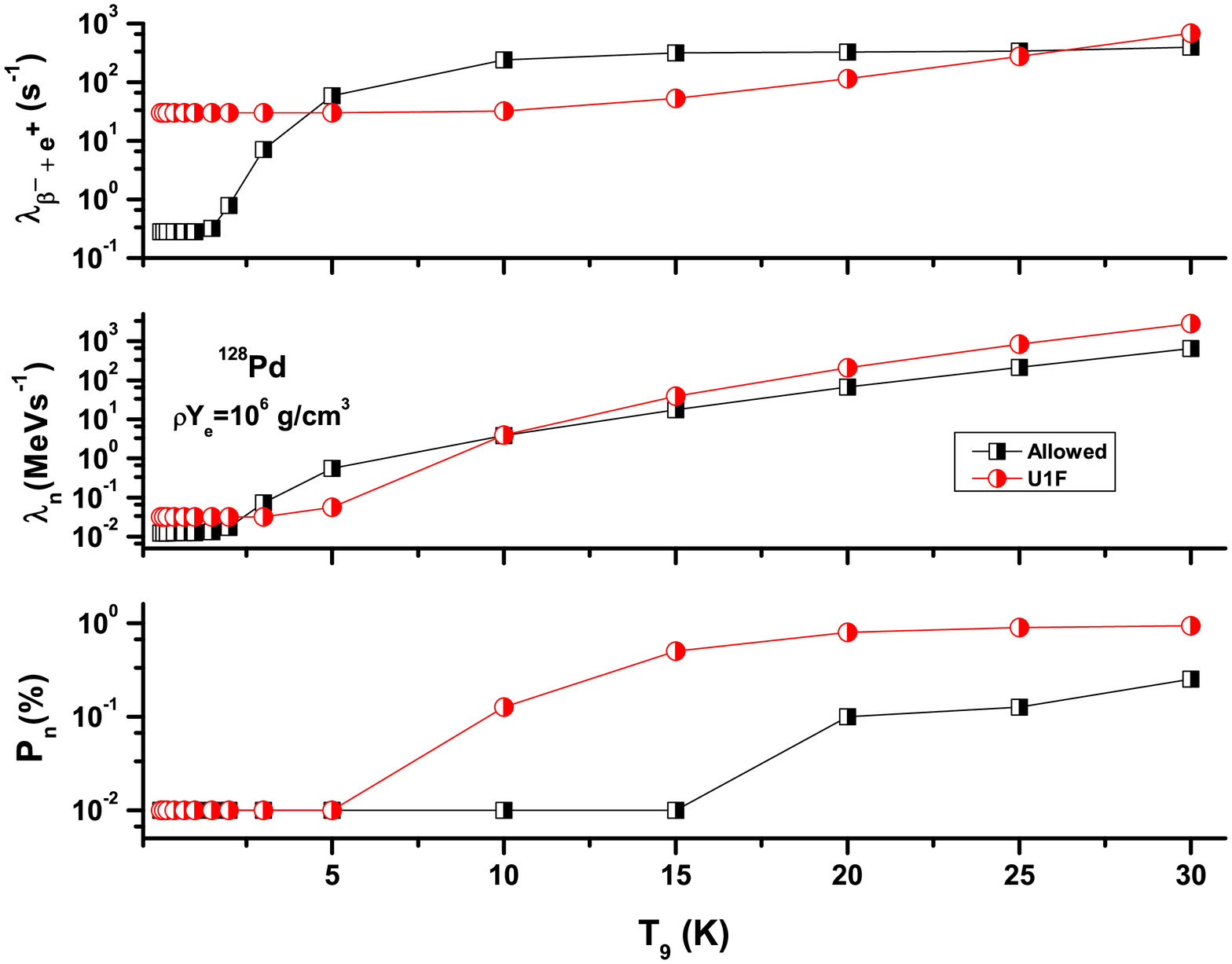}
\caption{\scriptsize Same as Fig.~\ref{figure1} but for
$^{128}$Pd.}\label{figure10} \centering
\end{figure}

Moving on from terrestrial to stellar environment,  we investigate
the  electron emission ($\beta^{-}$) rates and (continuum) positron
capture ($e^{+}$) rates for density range (10 - 10$^{11}$
g.cm$^{-3}$) and temperature range (0.01 $\leq$ T$_{9}$ $\leq$ 30,
where T$_{9}$ gives core temperature in units of GK), for our
selected ten $r$-process waiting point nuclei.
Figs.~\ref{figure1}-~\ref{figure10} show the calculated weak rates
for the ten selected nuclei. Each figure consists of three panels.
The upper panel shows the calculated sum of positron capture and
electron emission rates in stellar environment as a function of core
temperature.  It is to be noted that all parent excited states are
contributing to the calculated ($\beta^{-}$ and $e^{+}$) rate
calculation (see Eq.~(\ref{lb})). The  middle panel depicts the
calculated energy rates of $\beta$-delayed neutron in units of
$MeV.s^{-1}$. The bottom panel shows the calculated $\beta$-delayed
neutron emission probabilities. Within the Q$_{\beta}$ window, the
$\beta$-delayed neutron emission probabilities ($P$$_{n}$) are
required for the description of $\beta$ strength functions and
neutron separation energies. In all panels we show the allowed GT
and U1F contributions separately. All weak rates were calculated at
a fixed stellar density of 10$^{6}$ g.cm$^{-3}$ (simulating an
intermediate value of core density under stellar conditions).

For the intermediate density, the allowed rates in upper panel of
Figs.~\ref{figure1}-~\ref{figure3}, are up to an order of magnitude
bigger than the U1F rates at low stellar temperatures for $^{76}$Fe,
$^{79}$Cu and $^{81}$Ga, respectively. It is further noted that with
the increase of core temperature the U1F rates increase at a faster
pace and surpass the allowed rates at high $T_{9}$ values. At low
core temperatures, more $\beta$-delayed  neutrons are released due
to U1F transitions than due to GT transitions. Accordingly, at low
temperatures the energy rates of $\beta$-delayed neutron, due to U1F
transitions, is factor 2, factor 3 and up to an order of magnitude
bigger for $^{76}$Fe, $^{79}$Cu and $^{81}$Ga, respectively. The
energy rates due to U1F transitions are more than an order of
magnitude bigger at high stellar temperatures. The corresponding
emission probabilities due to U1F transitions are also appreciably
greater as can be seen in bottom panel of
Figs.~\ref{figure1}-~\ref{figure3}. At $T_{9}$ = 30 it is almost
certain that $\beta$-delayed neutrons would be emitted, both due to
allowed GT and U1F transitions.

Figs.~\ref{figure4}-~\ref{figure6} show similar results for the
waiting point nuclei $^{77}$Co, $^{78}$Ni and $^{80}$Zn,
respectively. Here the allowed $\beta$-decay rates are factor 3 -- 4
bigger than the corresponding U1F rates at low temperatures. At high
stellar temperatures the U1F rates supersede the GT rates by more
than an order of magnitude. The energy rates due to U1F transitions
are an order of magnitude bigger at low temperatures and even bigger
at high $T_{9}$ values.

Moving on to $N$ = 82 waiting point nuclei, Fig.~\ref{figure7} shows
the pn-QRPA calculated weak rates for $^{127}$Rh. Here the allowed
$\beta$-decay rates are factor 2 bigger than U1F rates at low
temperatures and more than order of magnitude bigger at high
temperatures. There is no crossing over between GT and U1F
$\beta$-decay rates as witnessed in previous figures
(Figs.~\ref{figure1}-~\ref{figure6}). However this crossing over of
rates is seen in the case of energy rates where U1F rates is an
order of magnitude bigger at low temperatures and factor 3 -- 5
smaller at high stellar temperatures.

The weak rate calculations for remaining three $N$ = 82 $r$-process
waiting point nuclei, $^{125}$Tc, $^{126}$Ru and $^{128}$Pd are
presented in Figs.~\ref{figure8}-~\ref{figure10}, respectively. The
$\beta$-decay rates due to U1F transitions are factor 3 bigger than
due to allowed transitions for $^{125}$Tc at low stellar
temperatures (Fig.~\ref{figure8}). At high temperatures the allowed
and U1F $\beta$-decay rates compete well. The middle panel shows
that the energy rates due to U1F rates are an order of magnitude
bigger at low temperatures and approach the allowed GT energy rates
at high temperatures. Upper panels of Fig.~\ref{figure9} and
Fig.~\ref{figure10} show that the U1F $\beta$-decay rates are an
order of magnitude bigger at low temperatures. The middle panels
show that UIF energy rates are factor 2 -- 6 bigger than allowed
energy rates at low and high $T_{9}$ values.

At low stellar temperatures, positron capture rates may safely be
neglected in comparison to $\beta$-decay rates. Only at high core
temperatures ($kT >$ 1 MeV), positron appears via $e^{-}$-$e^{+}$
pair creation. Positron capture rates becomes at par with
$\beta$-decay rates at T$_{9}$ = 30 (in fact for $^{81}$Ga they are
an order of magnitude bigger than $\beta$-decay rates). In general
the weak rates are product of phase space and reduced transition
probabilities (directly linked with strength distribution
functions).  The reason for the behavior of pn-QRPA calculated weak
rates depicted in Figs.~\ref{figure1}-~\ref{figure10} may be traced
to the allowed and U1F strength distributions and phase space
calculations which we discuss next.

\begin{figure}[t!]
\begin{center}
  \begin{tabular}{cc}
    \includegraphics[scale=0.30]{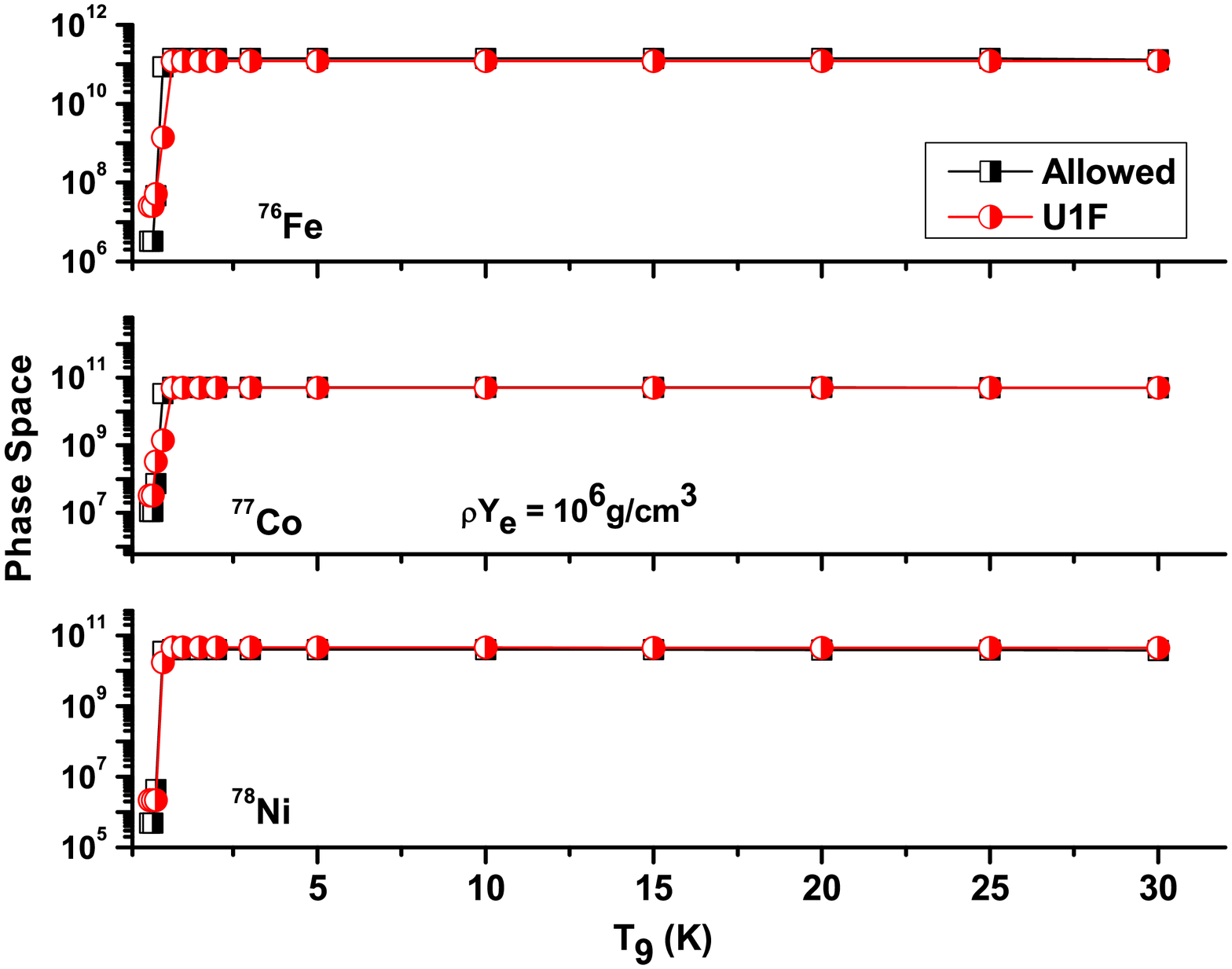} &
    \includegraphics[scale=0.30]{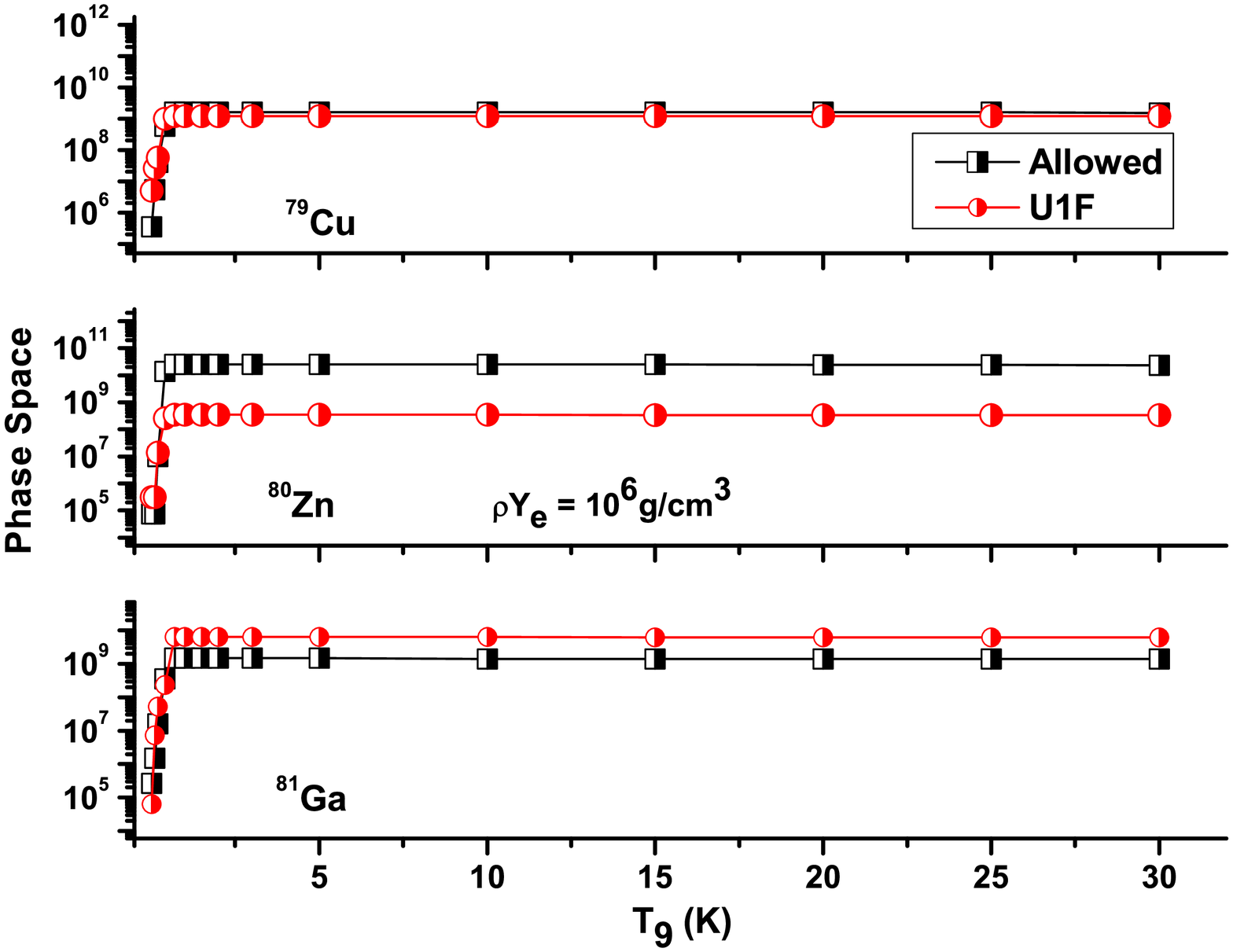}\\
\end{tabular}
\caption{Calculated phase space for allowed (GT) and unique
first-forbidden (U1F) $\beta$-decay for $N$ = 50 waiting point
nuclei as a function of stellar temperature at stellar density of
10$^{6}$g.cm$^{-3}$.}\label{figure11}
\end{center}
\end{figure}

\begin{figure}[t!]
\begin{center}
  \begin{tabular}{cc}
    \includegraphics[scale=0.30]{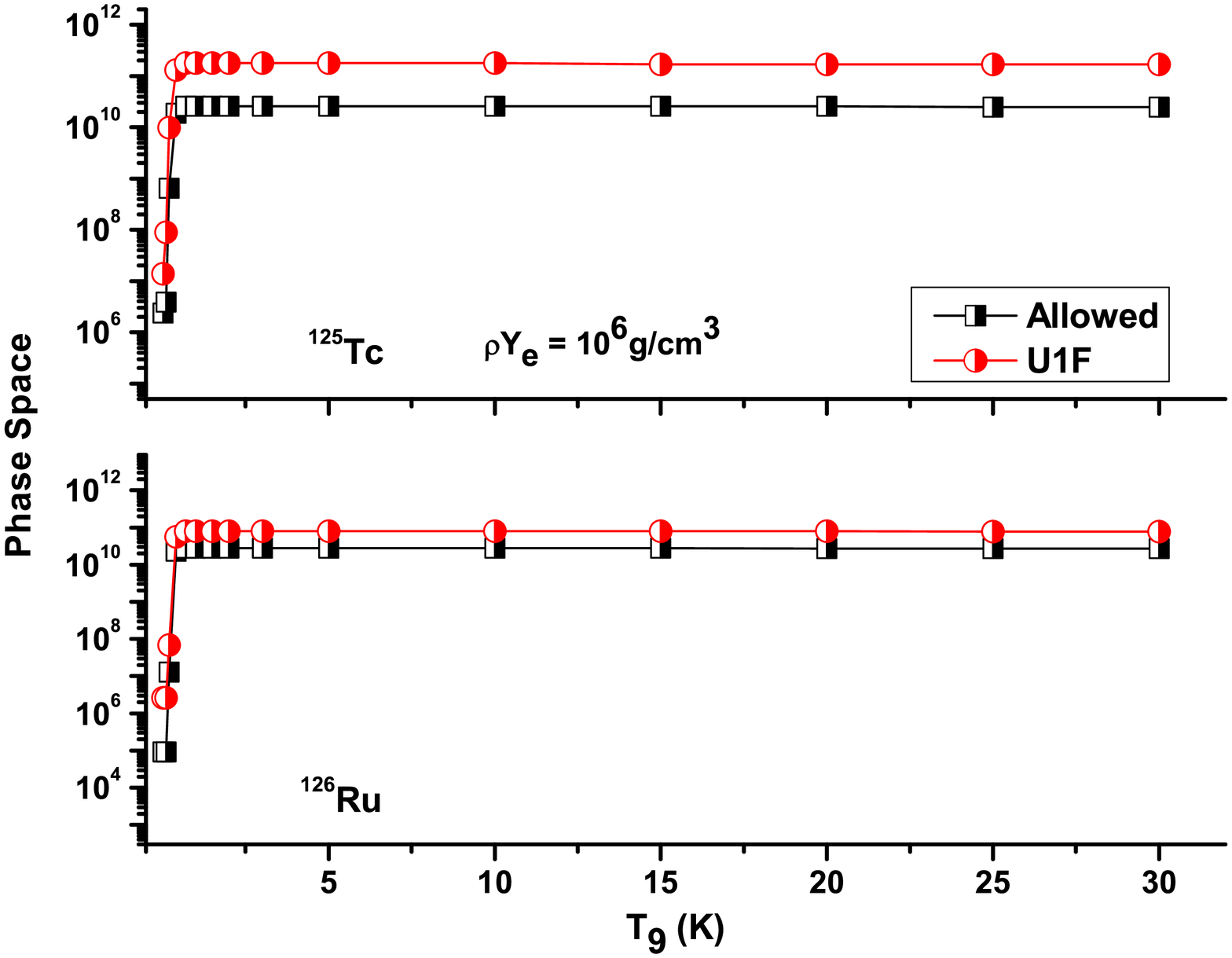} &
    \includegraphics[scale=0.30]{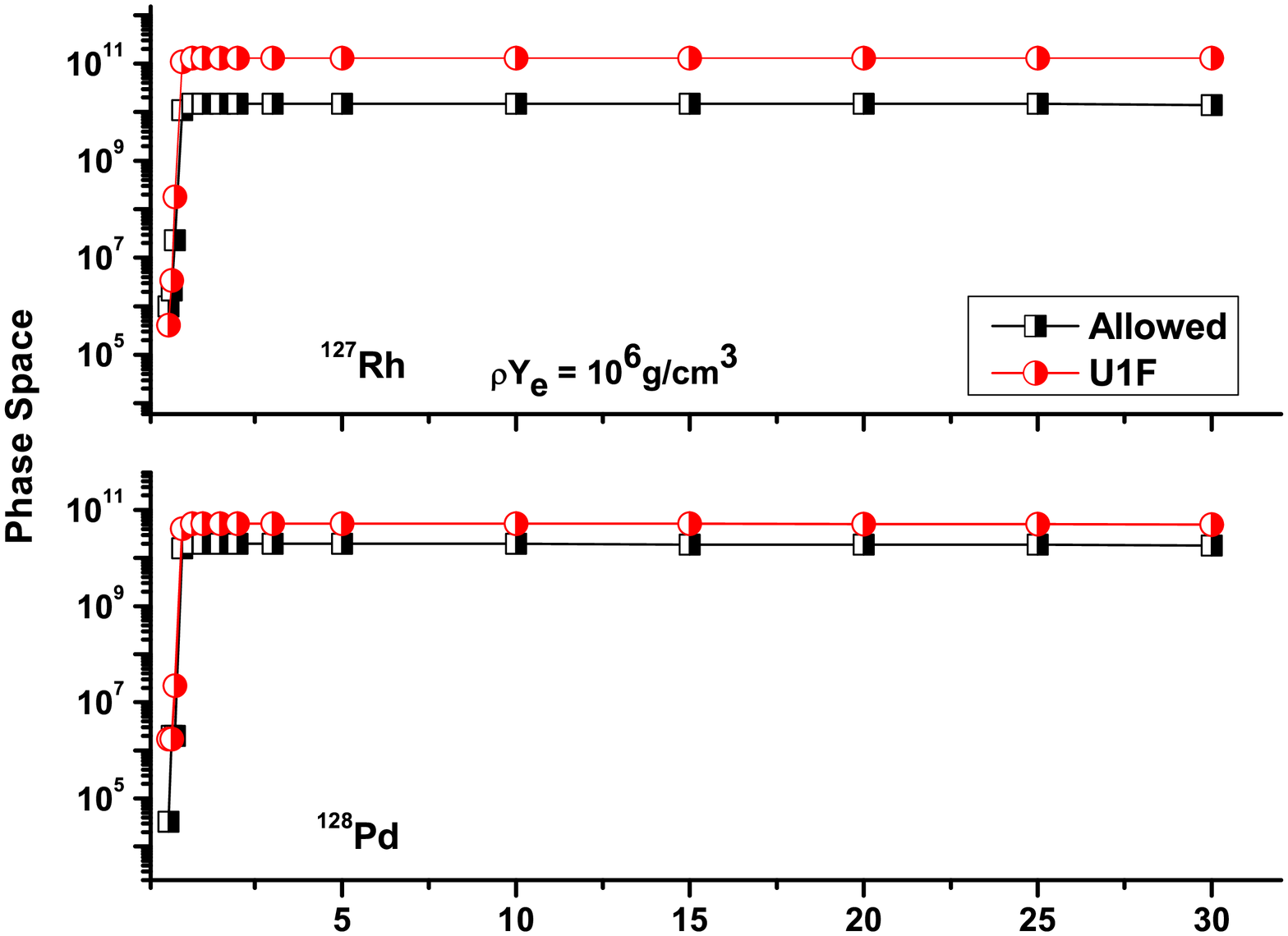}\\
\end{tabular}
\caption{Same as Fig.~\ref{figure11} but for $N$ = 82 waiting point
nuclei.}\label{figure12}
\end{center}
\end{figure}

In Fig.~\ref{figure11} and Fig.~\ref{figure12}, we show the phase
space calculation for allowed and U1F transitions as a function of
core temperature for  $N=50$ and  $N$ = 82 waiting point nuclei,
respectively,  at stellar density of 10$^{6}$ g.cm$^{-3}$. We chose
the same density at which we showed the calculation of weak rates
earlier.

The  phase space calculation for $N$ = 50 nuclei in
Fig.~\ref{figure11} displays certain distinctive features. At low
stellar temperatures, the U1F phase space is bigger by as much as an
order of magnitude compared to the allowed phase space. $^{81}$Ga is
an exception where the allowed phase space is bigger roughly by an
order of magnitude at $T_{9}$= 0.01. The phase space initially
increases at a fast pace till the core temperature approaches
$T_{9}$= 1. Beyond this temperature the phase space remains almost
constant till $T_{9}$= 30. At high temperatures the two phase spaces
are roughly the same for $^{76}$Fe, $^{77}$Co, $^{78}$Ni and
$^{79}$Cu. Allowed phase space is bigger (smaller) than U1F phase
space at high temperatures for $^{80}$Zn ($^{81}$Ga).

Fig.~\ref{figure12} shows few similarity of phase space calculation
for $N$ = 82 waiting point nuclei with the $N$ = 50 case. Once again
the allowed phase space is orders of magnitude smaller than the U1F
phase space at low temperatures (with the exception of $^{127}$Rh).
The rate of increase of phase space with temperature is rapid till
$T_{9}$= 1 and almost none beyond this temperature. In all four
cases we note that the U1F phase space is bigger by as much as one
order of magnitude at all temperatures (the only exception being
$^{127}$Rh at $T_{9}$= 0.01). This is one reason why U1F transitions
contribute significantly to the total weak rates for $N$ = 82
waiting point nuclei.

\begin{figure}[t!]
\begin{center}
  \begin{tabular}{|c|c|}
    \hline
    \includegraphics[scale=0.25]{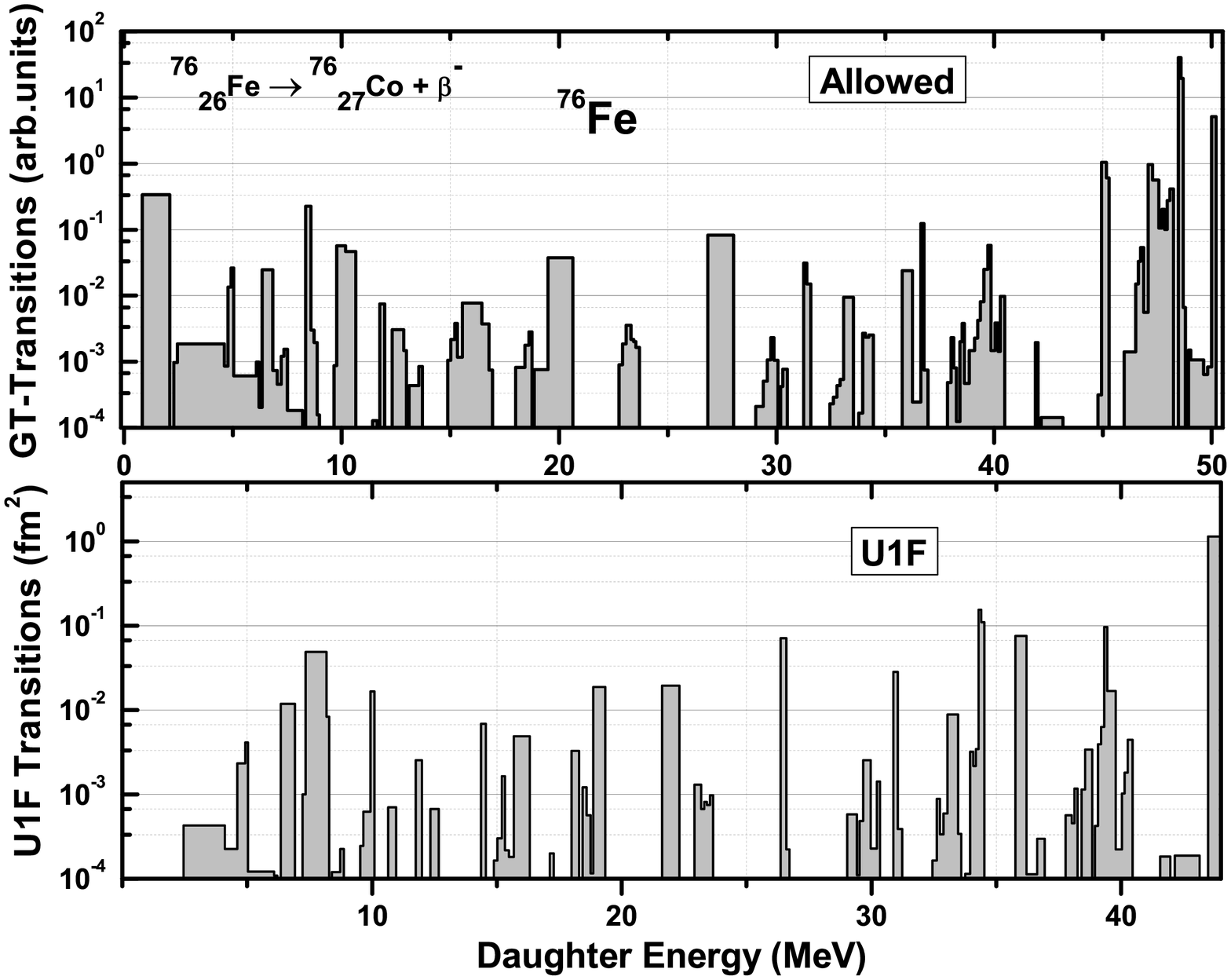} &
    \includegraphics[scale=0.25]{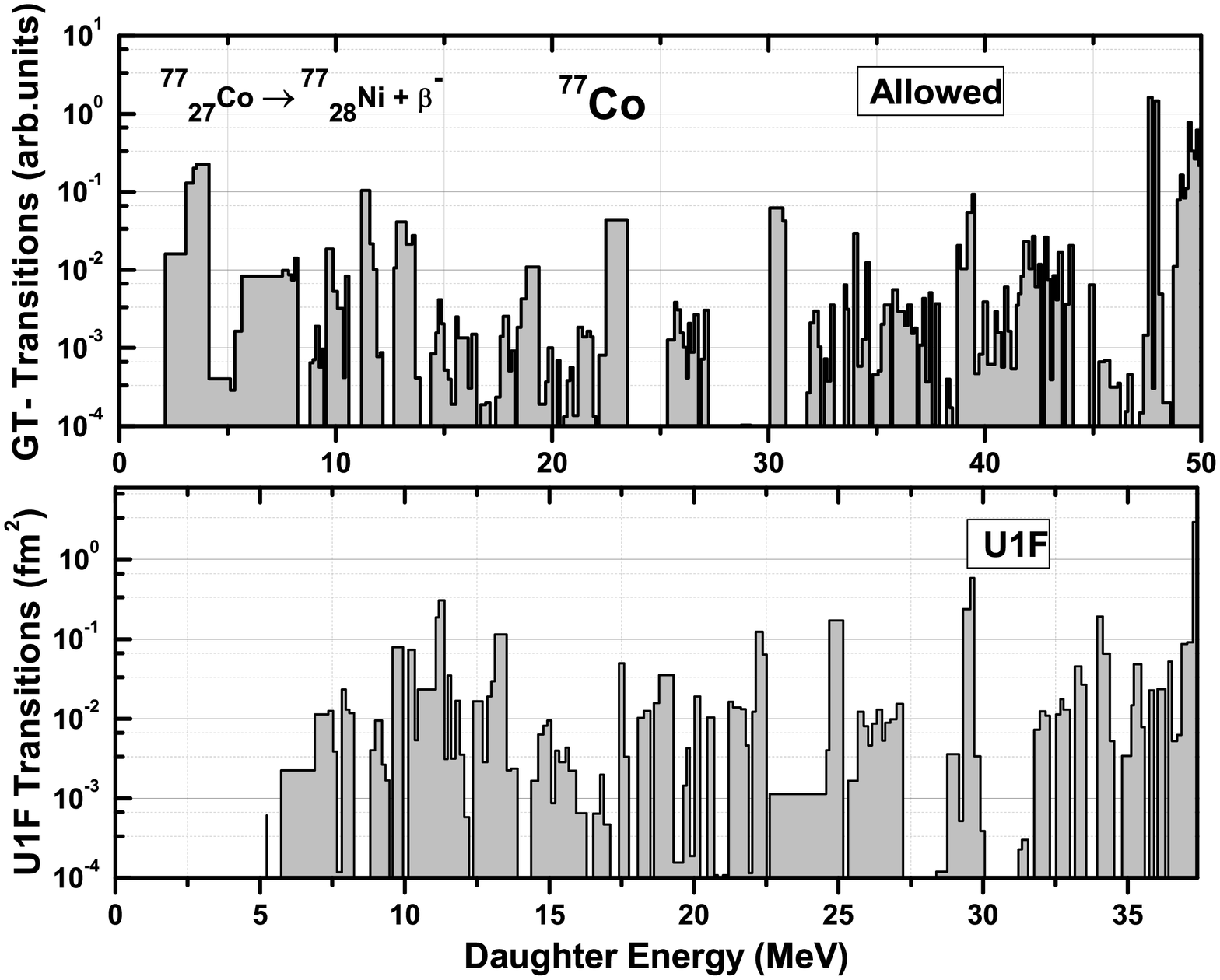}\\
    \hline
    \includegraphics[scale=0.25]{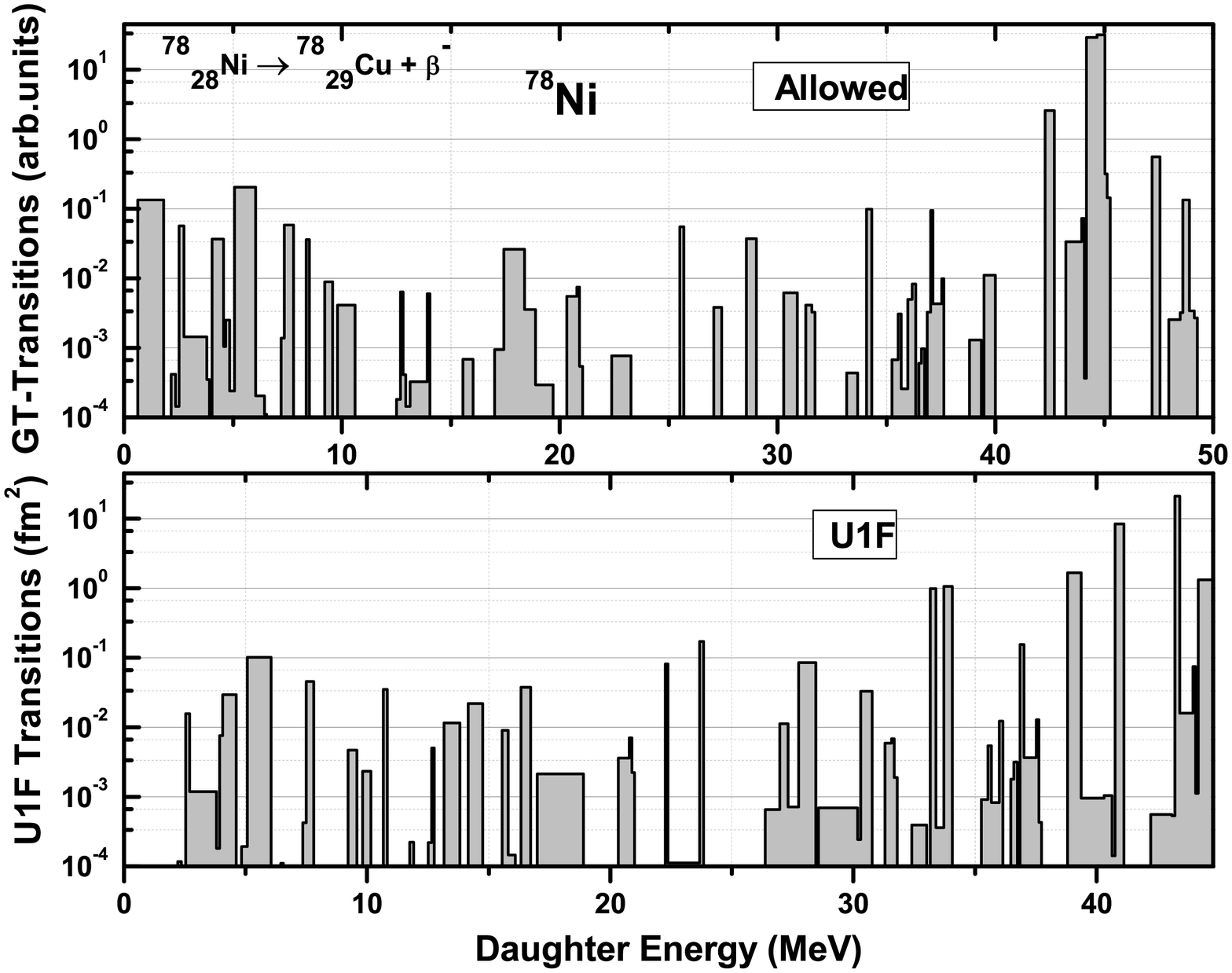} &
    \includegraphics[scale=0.25]{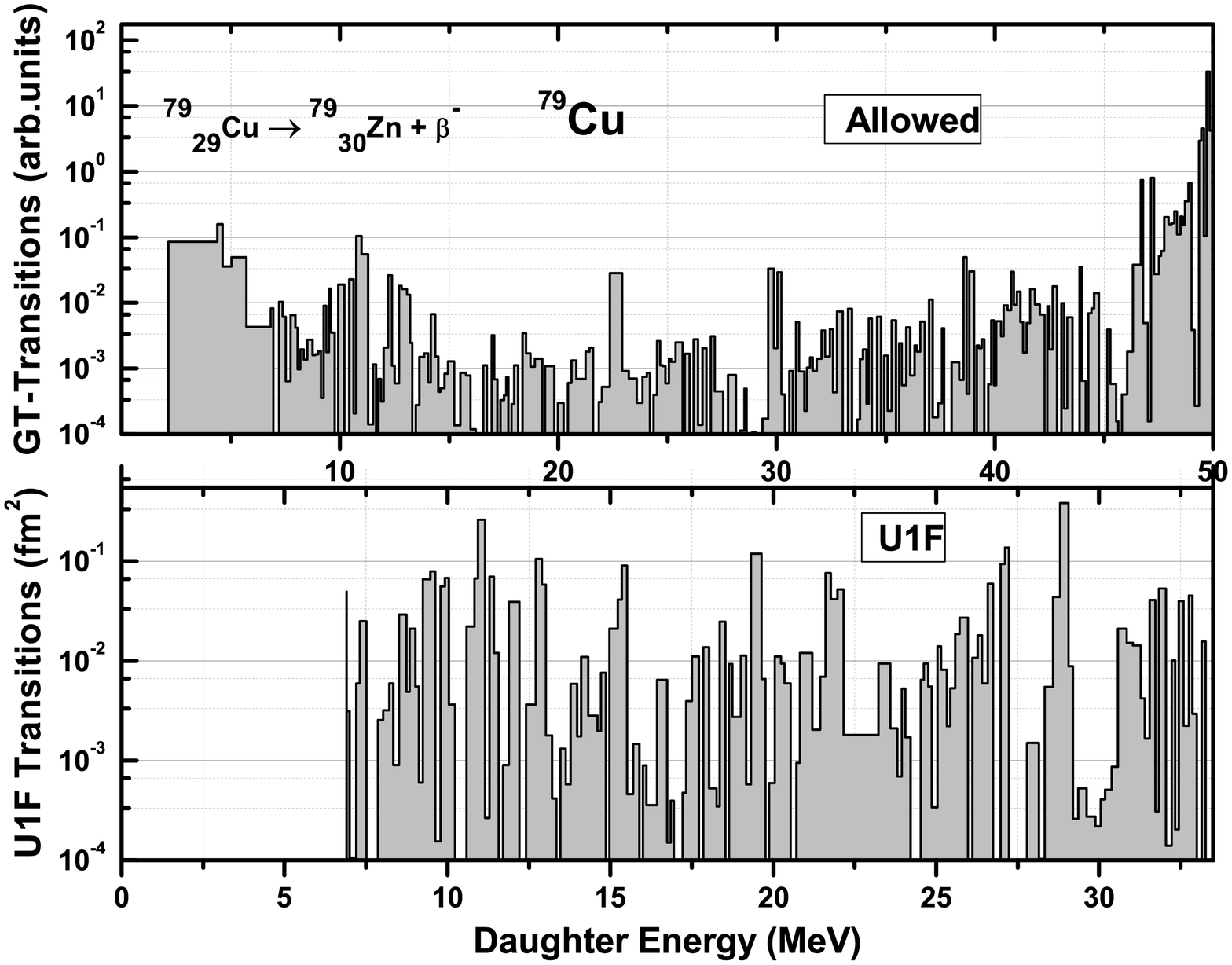}\\
    \hline
    \includegraphics[scale=0.25]{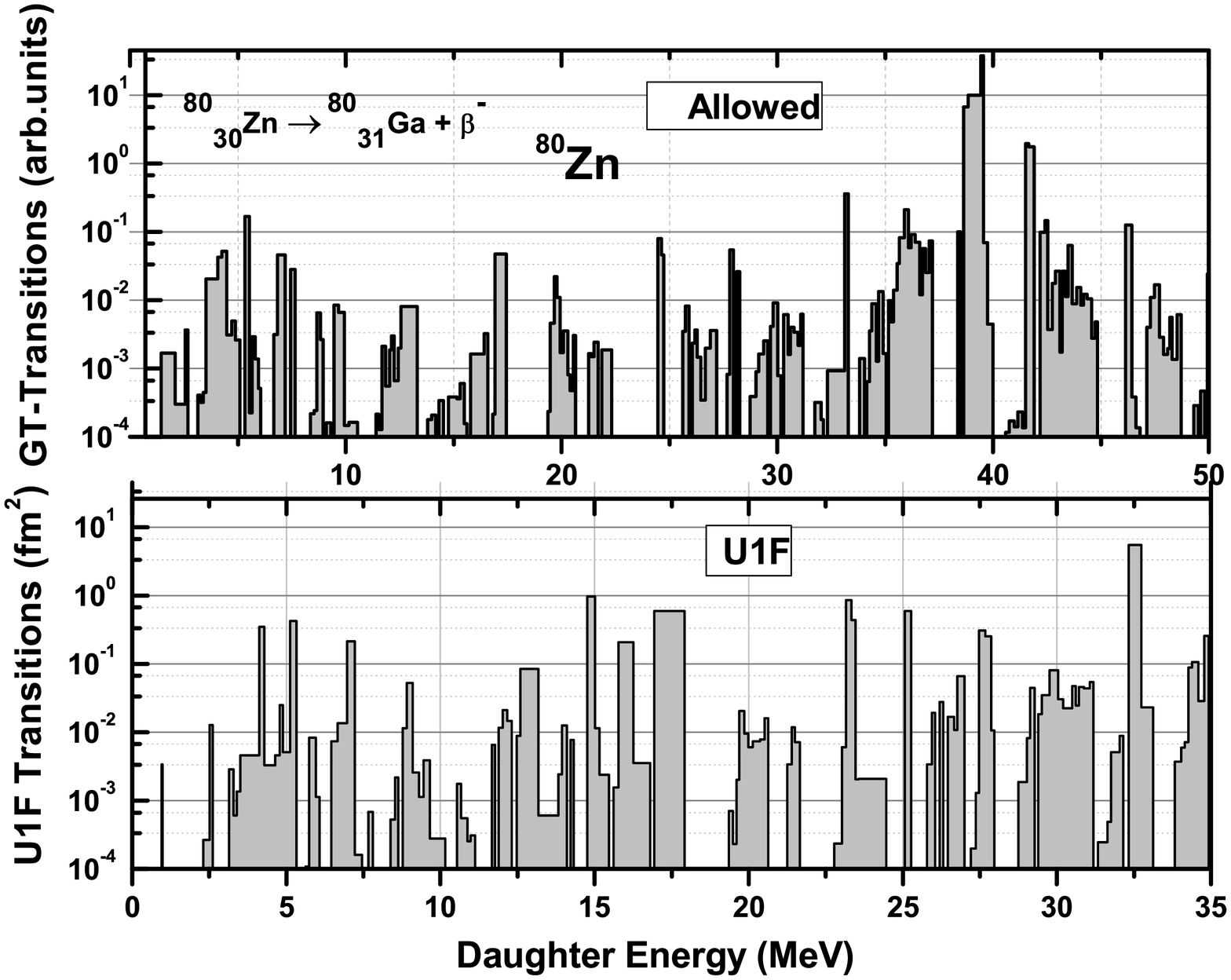} &
    \includegraphics[scale=0.25]{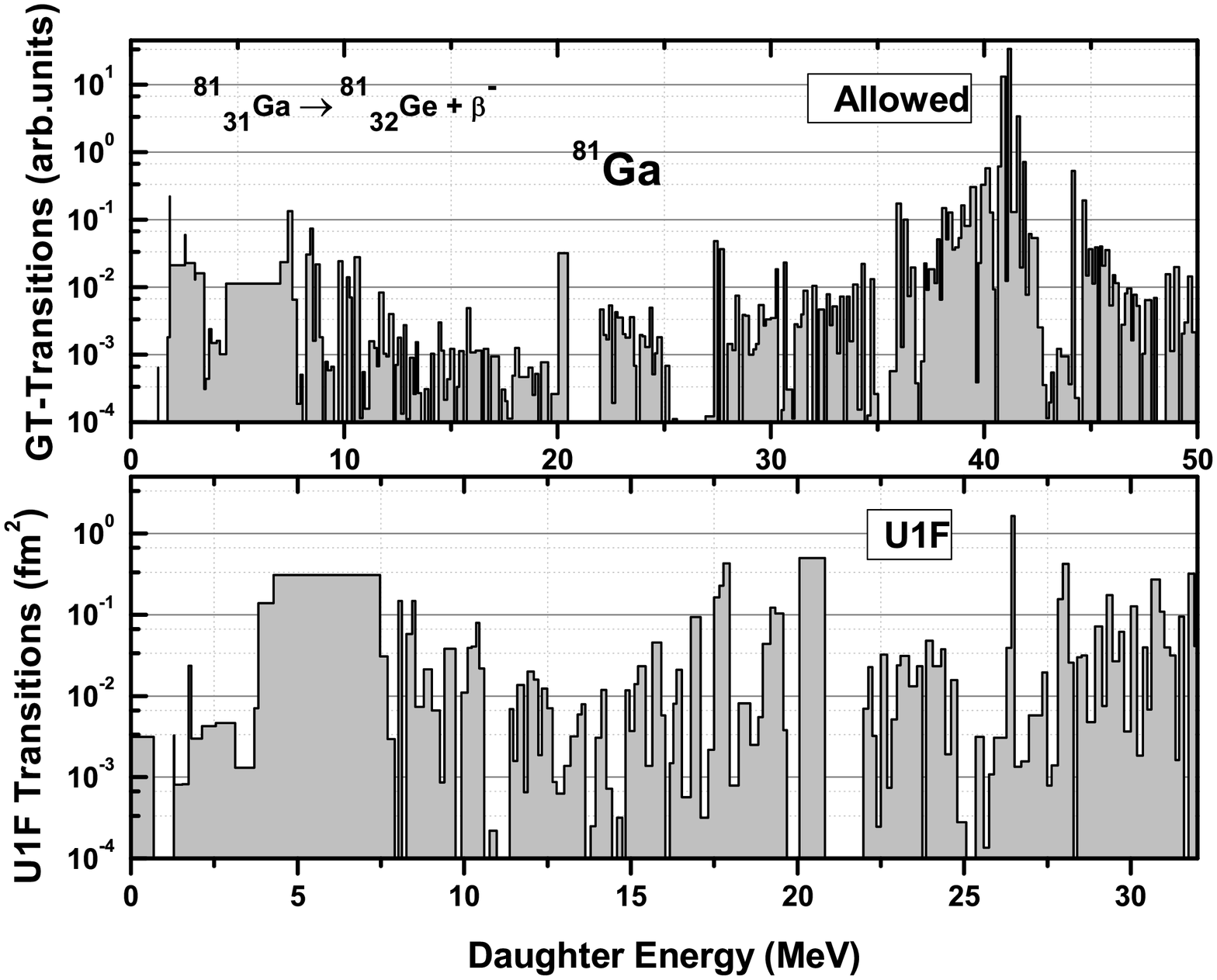}\\
    \hline
\end{tabular}
\caption{Allowed and unique-first forbidden (U1F) $\beta$-decay
transitions for $^{76}$Fe, $^{77}$Co, $^{78}$Ni, $^{79}$Cu,
$^{80}$Zn and $^{81}$Ga as a function of daughter excitation energy
calculated using the pn-QRPA model.}\label{figure13}
\end{center}
\end{figure}

\begin{figure}[t!]
\begin{center}
  \begin{tabular}{|c|c|}
    \hline
    \includegraphics[scale=0.25]{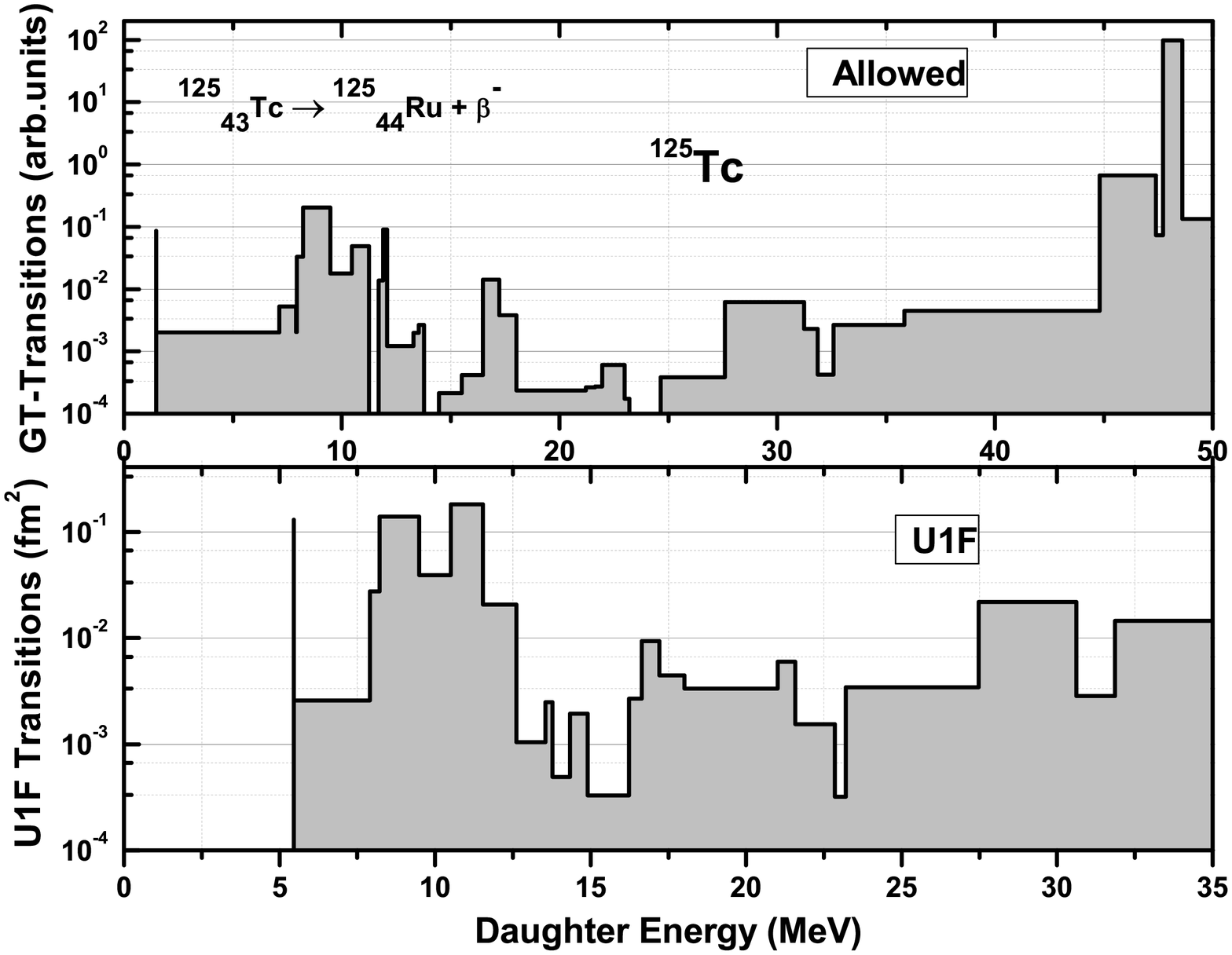} &
    \includegraphics[scale=0.25]{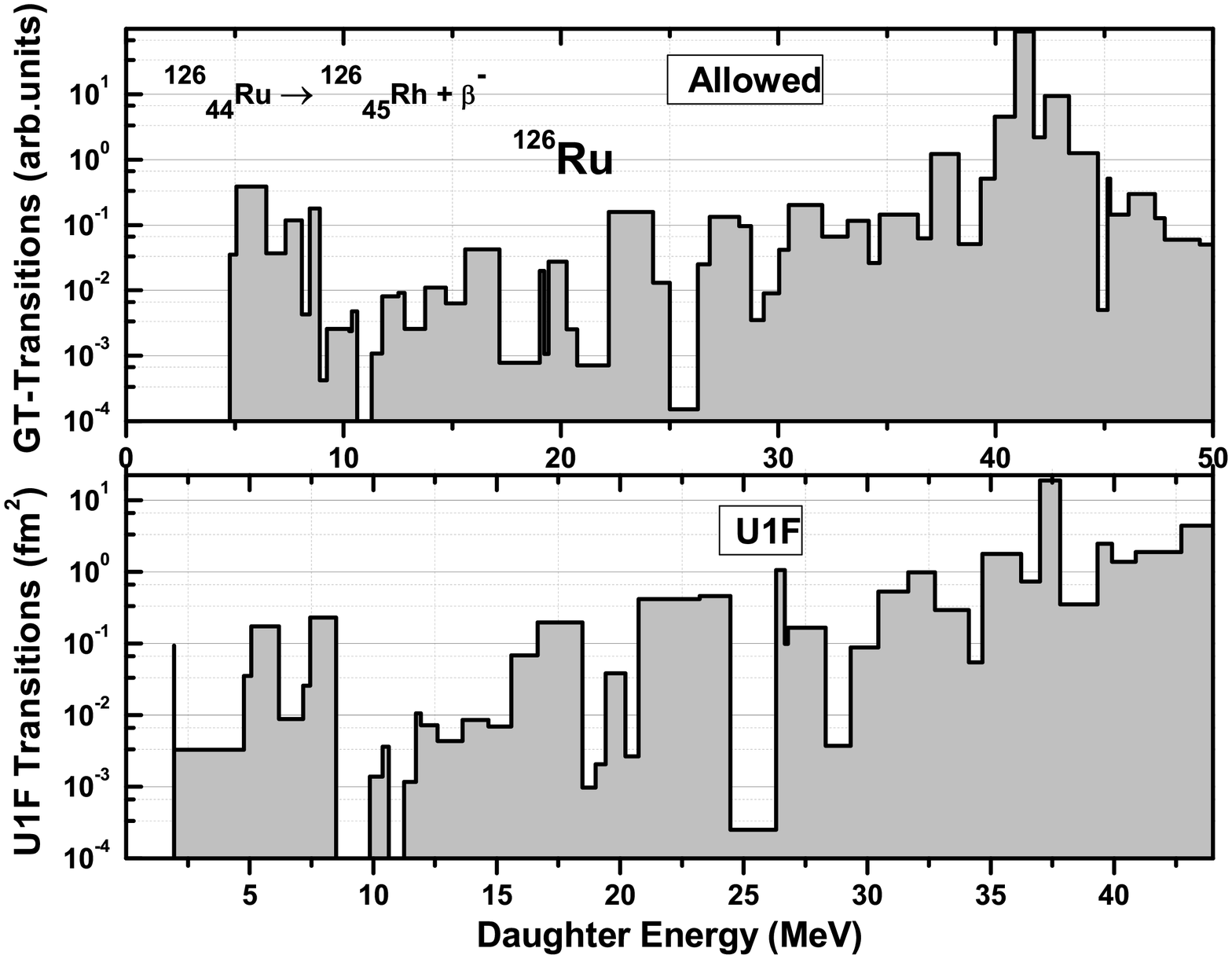}\\
    \hline
    \includegraphics[scale=0.25]{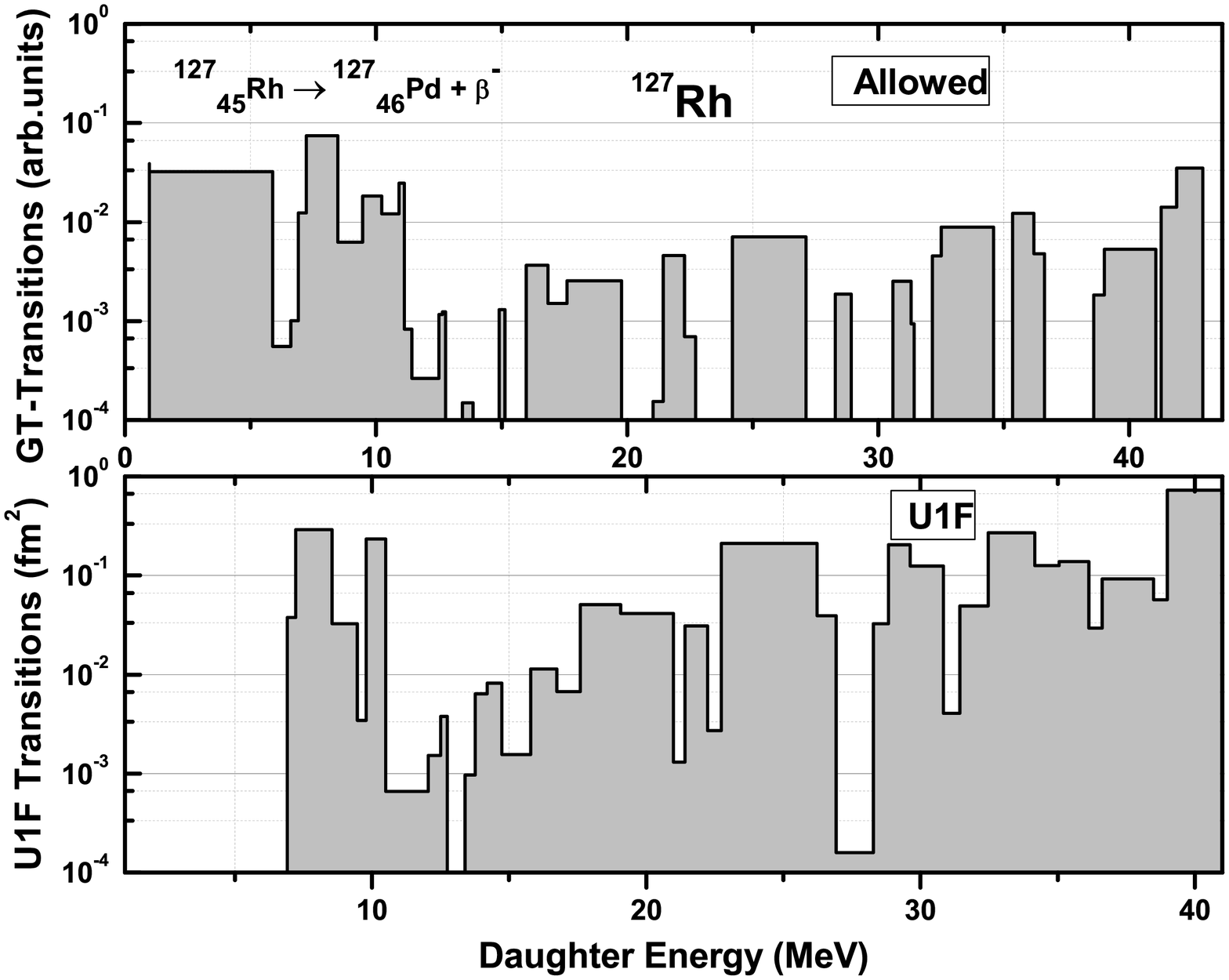} &
    \includegraphics[scale=0.25]{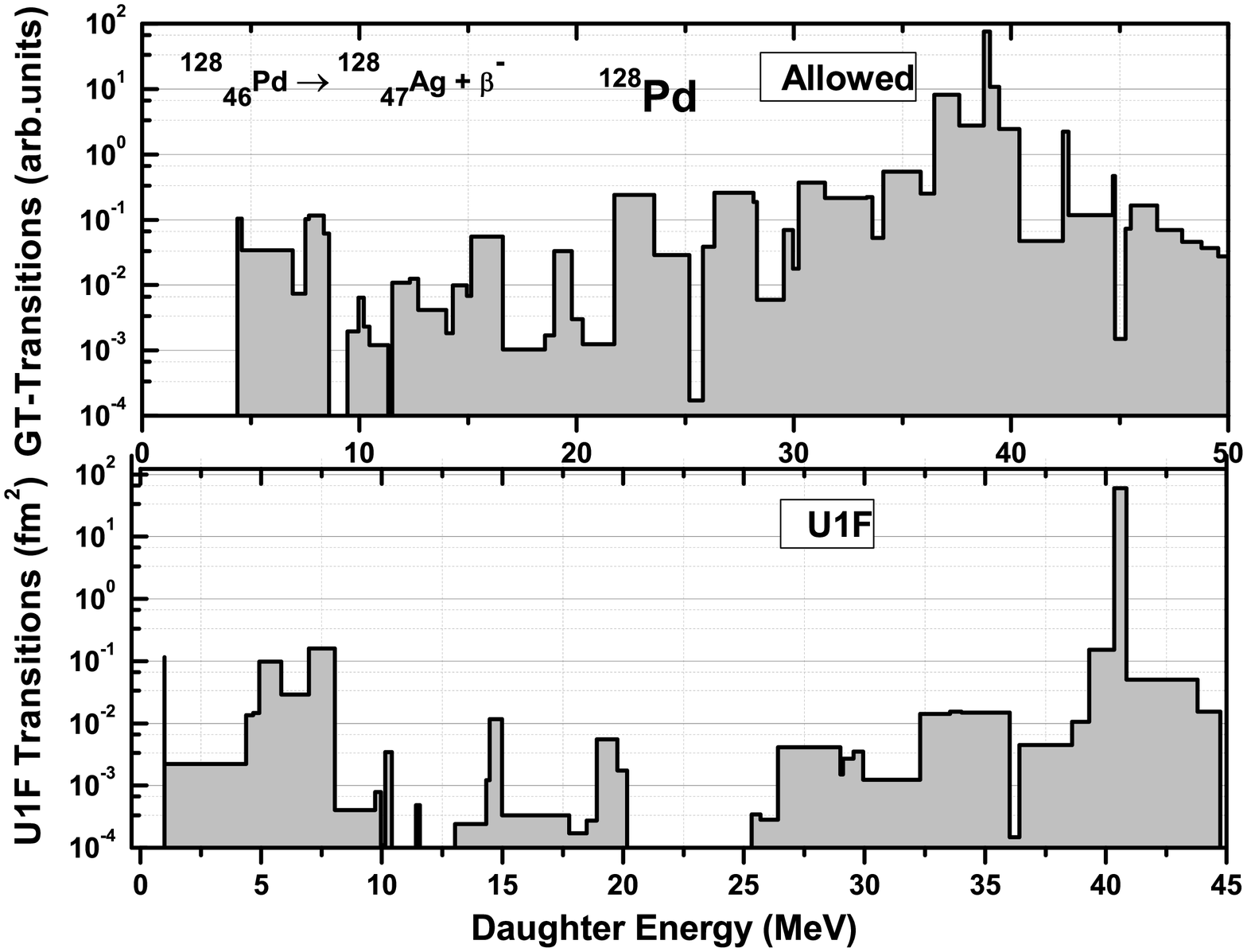}\\
    \hline
\end{tabular}
\caption{Same as Fig.~\ref{figure13} but for $^{125}$Tc, $^{126}$Ru,
$^{127}$Rh and $^{128}$Pd.}\label{figure14}
\end{center}
\end{figure}

The skyscrapers for the calculated charge-changing strength
distribution along $\beta$-decay direction for the ten waiting point
nuclei are shown in Fig.~\ref{figure13} and Fig.~\ref{figure14}. For
each nucleus, the bottom panel shows the U1F transitions and the
upper panel the allowed GT transitions, respectively. We again
mention that all our calculated strength were quenched by a factor
of $f_{q}^{2}$ = (0.55)$^{2}$ \cite{Nab07}. It may be noted that the
pn-QRPA model calculates high lying allowed GT transitions for all
nuclei. This is also verified by the large values calculated values
of centroid shown in Table~\ref{tab4}. The U1F transitions were
calculated to relatively lower excitation energies in daughter. This
trend was also noted in the calculation of \cite{Mol03}. It is noted
in Fig.~\ref{figure13} that U1F transitions are comparable in
magnitude with the allowed GT transitions for $^{78}$Ni and
$^{80}$Zn. This is the reason that the terrestrial half-life is
reduced by $\sim$ 50$\%$ when U1F transitions were incorporated in
pn-QRPA calculation (see Table~\ref{tab1}). We note  significant
contribution from U1F transitions for the $N$ = 82 cases
(Fig.~\ref{figure14}). For the even-even cases, $^{126}$Ru and
$^{128}$Pd, this substantial U1F contribution resulted in more than
95$\%$ reduction in calculated terrestrial half-lives (see
Table~\ref{tab1}). For the case of $^{127}$Rh, once again the U1F
contribution is very significant but from Table~\ref{tab1} the
half-life is reduced only by 32$\%$. The reason for this is traced
back to phase space calculation where it is seen from
Fig.~\ref{figure12} that at T$_{9}$ = 0.01 (at this low temperature,
stellar phase space would very much mimic the terrestrial phase
space) the U1F phase space is roughly an order of magnitude smaller
than the allowed phase space.

\section{Summary and Conclusions}
For the first time we present the allowed GT and U1F weak rates of
$N$ = 50 and $N$ = 82 waiting point nuclei in stellar environment
using the deformed pn-QRPA model. We quenched our calculated
charge-changing transition by a quenching factor of $f_{q}^{2}$ =
(0.55)$^{2}$. The calculated charge-changing strength distributions,
phase space and weak rate calculations, separately for allowed and
U1F transitions, were presented for a total of ten $r$-process
waiting point nuclei. High lying centroids were computed for the
calculated allowed GT strength distributions. Our calculation
fulfilled the model-independent Ikeda sum rule, except for a couple
of odd-A cases. The pn-QRPA calculated half-lives, after
incorporation of U1F transitions, were in decent agrement with the
measured half-lives and at the same time were also suggestive of
incorporation of non-unique forbidden contributions which we plan to
take as a future assignment. It is hoped that the present study
would prove useful for a better and reliable simulation of
nucleosynthesis calculation.

We found substantial U1F contribution to the $\beta$-decay
half-lives for the $N$ = 82 waiting points. Except for $^{127}$Rh,
the calculated U1F stellar rates were orders of magnitude bigger
than allowed GT rates at high stellar temperatures approaching
T$_{9}$ = 30.

The neutrino-driven wind streaming out of the neutron star forming
at the center of a type II supernova has been shown to be one of
possible candidates for the cite of $r$-process. If $r$-process
happened in a neutron-rich environment, then the electron neutrino
capture could compete with the $\beta$-decay rates and is capable of
modifying the $r$-abundance distribution by subsequent $\nu$-induced
neutron spallation. We plan to calculate the charged-current
electron neutrino capture as a future assignment.

The weak rates for all ten waiting point nuclei, as a function of
stellar temperature and density, are available as ASCII files and
may be requested from the corresponding author.

\vspace{0.5in} \textbf{Acknowledgment}:  J.-U. Nabi would like to
acknowledge the support of the Higher Education Commission Pakistan
through the HEC Project No. 20-3099.

\end{document}